\documentclass[reprint,amsmath,amssymb,aps,prb,floatfix,longbibliography]{revtex4-2}
\usepackage{graphicx,bm}
\usepackage{xcolor}
\usepackage{amssymb} 
\usepackage{dsfont}
\usepackage{braket}
\graphicspath{{figs/}}
\makeatletter
\NewCommandCopy\orig@mathbb\mathbb 
\DeclareRobustCommand{\mathbb}[1]{\def\@mbbarg{#1}\def\@mbbone{1}%
  \ifx\@mbbarg\@mbbone\mathds{1}\else\orig@mathbb{#1}\fi}
\makeatother

\newcommand{\Ev}{\mathbb{E}}
\newcommand{\Tr}{\operatorname{Tr}}

\usepackage{hyperref}
\hypersetup{colorlinks=true, linkcolor=blue, citecolor=blue, urlcolor=blue}

\begin{document}

\title{High-Capacity Generalized Hopfield Networks}
\author{Victor Galitski}
\affiliation{Joint Quantum Institute, Department of Physics, University of Maryland,
College Park, MD 20742, USA}

\begin{abstract}
Generalized Hopfield networks are introduced where memories and neurons are continuous variables that lie on a Riemannian manifold. We explicitly focus on symmetric spaces associated with the special unitary groups $SU(d)$, and use both numerical and analytical (replica) techniques to demonstrate an almost order of magnitude enhancement in critical capacity  over the vector  networks  starting with $d=3$ and further rapidly growing with $d$. To circumvent the non-linear geometric constraints, we use a Lie algebraic method [following V.~Galitski, Phys. Rev.~A {\bf 84}, 012118 (2011)] to exactly describe the classical neural network in terms of linear algebra in an auxiliary Hilbert space. It is shown that in contrast to the traditional Hopfield networks, memory recall in  $SU(d)$ Hopfields corresponds to neuron alignment along a top eigenvector of a spiked  matrix, which is less susceptible to random matrix crosstalk than other models with continuous neuron variables. Physical platforms to realize $SU(d)$ Hopfields are briefly discussed and physical (in addition to algorithmic) recall mechanism is demonstrated, where memory recovery occurs naturally through generalized Landau-Lifshitz-Gilbert dynamics. To illustrate $SU(3)$ memory recall, we introduce a color (RGB) image encoding/decoding protocol and explicitly run image recovery on corrupted cues. Finally, we quantize the generalized Hopfields which are shown to reduce to Sachdev-Ye glassy type of models. Their many-body spectra generally feature two types of dark and memory bands, where the latter exhibits chaotic Wigner-Dyson level statistics that hides Hebbian data. 
\end{abstract}

\maketitle

\section{Introduction}
The input data into an $N$-neuron Hopfield network~\cite{Hopfield1982} is the Hebbian coupling matrix constructed out
of ${P}$ memories ${\bm \xi}_i^\mu$,
\begin{equation}
\label{J}
\hat{J}_{ij} = \frac{1}{N} \sum\limits_{\mu=1}^{P} {\bm \xi}_i^\mu \, ({\bm \xi}_j^\mu )^{\rm T}.
\end{equation}
Indices $i$ and $j$ label neurons, $\mu$ labels memories, and ${\rm T}$ denotes the transpose vector. For canonical binary networks no vector notation is necessary and
${\xi}^\mu \in \{\pm1\} = \mathbb{S}^0$; for phasor networks the neurons lie on a circle ${\bm \xi}_i^\mu \in \mathbb{S}^1$; and for vector Hopfields they  lie on a $2$-sphere ${\bm \xi}_i^\mu \in \mathbb{S}^2$. The energy functional is
\begin{equation}
\label{HE}
{\cal E} = - \frac{1}{2} \sum\limits_{i,j=1}^N {\bm s}_i^{\rm T} \hat{J}_{ij} {\bm s}_j =
- \frac{N}{2} \sum\limits_{\mu=1}^{P} O^2_\mu,
\end{equation}
where the spins ${\bm s}_i$ belong to the same sphere as the memories and the overlap
$O_\mu= \frac{1}{N} \sum\limits_{i=1}^N {\bm s}_i \cdot {\bm \xi}_i^\mu$ measures the alignment between the state of the
network and memory $\mu$.  Maximizing $O_\mu$ corresponds to the perfect alignment ${\bm s}_i = {\bm \xi}_i^\mu$, while
$O_\mu\sim 1/\sqrt{N}$ corresponds to a random state unrelated to the memory, $\mu$. We are typically interested in the large-$N$
limit.

For sufficiently small $\alpha=P/N$, a slightly corrupted memory ($O_\mu \lesssim 1$) can be restored as follows.
Pick spins one by one and align them along the mean field produced by the other spins. This leads to a monotonic
decrease of the energy until it reaches a fixed point - the state of the system corresponding or close to the desired memory.

However, if the storage load exceeds a critical value $\alpha_c$, the noisy crosstalk between neurons dominates the
signal and recovery is not guaranteed. The critical capacity is $\alpha_c(\mathbb{S}^0)\approx 0.138$ for the binary
network~\cite{AGS1985,AGS1987}, $\alpha_c(\mathbb{S}^1)\approx 0.07$ for the phasor~\cite{Noest1988,NoestPRA1988}, and
$\alpha_c(\mathbb{S}^2)\approx 0.05$ for the vector (Heisenberg) network~\cite{Barney2026}; more generally
$\alpha_c(\mathbb{S}^{n-1})\approx 4/(27n)$ for the higher-dimensional $n$-sphere~\cite{Nicoletti2025}. The capacity thus
\emph{decreases} with the sphere dimension, because a continuous vector drifts off a stored direction more easily
with increasing the number of continuous degrees of freedom. This is in contrast to discrete multi-state Hopfields~\cite{Kanter1988,KropffTreves2005} and the higher-order (``dense'') associative memories~\cite{KrotovHopfield2016,Demircigil2017}.

These results might suggest that further increasing the complexity of the ``neuron manifold'' is a suboptimal route to higher
capacity. This work shows the opposite: a large class of manifolds with richer {geometry} and topology than a sphere support a qualitatively new recall mechanism and much higher capacity. As shown below, the new mechanism aligns the {eigenvector} of a matrix mean-field object - memory kernel - which is less sensitive to random-matrix crosstalk than vector alignment. This is demonstrated both numerically and using analytical methods. For large $d$, the replica analysis predicts $\alpha_c \gg 1$.

{This paper is structured as follows. Section~\ref{sec:symmetric} defines Hopfield networks on symmetric spaces and constructs the $SU(d)$ family on $\mathbb{C}P^{d-1}$. Section~\ref{sec:sud} derives the recall rule: alignment of a neuron along the top eigenvector of its memory kernel. Section~\ref{sec:image} introduces the color-image encoding and visualizes recall restoring a corrupted fragment of a photograph by running the update rule. It also estimates the critical capacity using numerical analysis using Haar random memories/images. Section~\ref{sec:capacity} explains the enhanced capacity, through both a qualitative spiked random-matrix picture and quantitatively through the replica analysis. Section~\ref{sec:quantum} quantizes the generalized Hopfields. Section~\ref{sec:xy} examines the many-body spectra of the minimal quantum Hopfield network, which is shown to exhibit Wigner-Dyson level statistics. Section~\ref{sec:recall} constructs the generalized Landau-Lifshitz-Gilbert  dynamics - applicable to both classical and mean-field quantum Hopfields - and explicitly demonstrates memory recall through real-time evolution, winner-takes-it-all dynamics, and a shadow phenomenon for close memories.}

\section{Hopfield network on symmetric spaces}
\label{sec:symmetric}

Let ${\cal M}$ be a compact Riemannian manifold. $N$ neurons and $P$ memories live on this manifold $s_i, \xi_i^\mu \in {\cal M}$. We shall need the notion of a scalar product between the neurons and memories. For this, we consider a smooth embedding of the manifold in the Euclidean space: ${\cal M} \subset \mathbb{R}^D$. Choose a system of coordinates in $\mathbb{R}^D$ and we can define explicit {$D$-dimensional} vectors ${\bf s}_i=(s_i^1, s_i^2,\ldots, s_i^D)$ and ${\bm \xi}_i=(\xi_i^1, \xi_i^2,\ldots, \xi_i^D)$ and their dot-products. This defines a generalized Hopfield network on ${\cal M}$, which can be written in  exactly the same  way as in Eqs.~(\ref{J}) and (\ref{HE}). The only difference is that both memories and neurons are free to explore the more general manifold ${\cal M}$. This structure is of course embedding-dependent, but the group-theoretic analysis of symmetric spaces below will provide a natural choice of the embedding and make all the algebraic structures and the definition of the generalized Hopfield unique.

We now focus on specific manifolds - symmetric spaces - associated with a symmetry group, $G$, and a subgroup, $H$: ${\cal M} = G/H$. Note that all standard Hopfield networks can be presented in such form because an $n$-dimensional sphere can be written in terms of the rotation group $\mathbb{S}^{n} = O(n+1)/O(n)$ (formally, even the binary case can be written this way $\{+1,-1\}=\mathbb{Z}_2=\mathbb{S}^0=O(1)/O(0)$ with $O(0)=\{e\}$ is a trivial single-element group, although of course this construction is unnecessary in isolation). As discussed, the spherical Hopfields are not promising from the perspective of increasing capacity. There are infinitely many choices of other manifolds  but for the sake of concreteness, we shall focus on the well-studied class of symmetric spaces associated with the special unitary group, $SU(d)$, and its defining matrix representation. I.e., the group elements are $d 
\times d$ unitary complex matrices, $\hat{g} \in G=SU(d)$, acting on $d$-dimensional complex vectors, $\ket{\psi}=(\psi_1,\psi_2,\ldots,\psi_d)^T \in \mathbb{C}^d$. The generators of the group are $D=(d^2-1)$ $d \times d$ Hermitian matrices, $\{ \hat{\lambda}_a \}_a^D/2$. Each group element can be written as
$\hat{g} = e^{i  \phi_a \hat{\lambda}_a/2} \in SU(d)$, where the sum over repeating indices is assumed.
The generators satisfy the canonical commutation relations
\begin{equation}
\label{cr}
[ \hat{\lambda}_a, \hat{\lambda}_b ] = 2i f_{ab}^c \hat{\lambda}_c,
\end{equation}
where $f_{ab}^c$ are known structure constants. For example, for $d=2$ $f_{ab}^c= \varepsilon_{abc}$ is the antisymmetric tensor and the generators are the Pauli matrices and for $d=3$, the generators are Gell-Mann matrices. For the sake of brevity, we shall refer to the generators as the Gell-Mann matrices for all $d$. Note that the construction is completely classical,  and we use the quantum-mechanical vocabulary solely for the sake of convenience (although, it makes the Hopfield quantization natural following author's Ref.~\cite{VG11}). To generate a manifold we use Perelomov construction~\cite{Perelomov1972} of the (``most classical'') coherent states that live on the generalized Bloch sphere, which is our ${\cal M}$ defined as follows. 

Fix a reference vector in $\mathbb{C}^d$, for example $\ket{\psi_0}=(1,0,\ldots,0)^T$. Define a maximal stability subgroup of $G=SU(d)$ as the set of all group elements that leaves the reference state intact modulo a $U(1)$ phase. I.e., 
\begin{equation}
\label{H}
H = \left\{ \hat{h}\in SU(d): \hat{h} \ket{\psi_0} = e^{i\phi} \ket{\psi_0}, \phi\in[0,2\pi) \right\}.
\end{equation}
It can be proven that $H = U(d-1)$ and hence the symmetric space is
\begin{equation}
\label{M}
{\cal M} = SU(d)/U(d-1) =\mathbb{C}P^{d-1},
\end{equation}
which is a $(d-1)$-dimensional complex projective space (of real dimension $2(d-1)$). Both conveniently and confusingly, a useful way to define and ``visualize'' this complex space is as a set of normalized $d$-dimensional ``qudits'' defined modulo a $U(1)$ phase. I.e., we can use the familiar ``quantum-mechanical'' notations to describe both memories and spins, e.g., 
$$
\ket{\xi_i^\mu} = \left( \xi_{i1}^\mu, \xi_{i2}^\mu, \ldots,\xi_{id}^\mu \right)^T
$$
where $\braket{\xi_{i}^\mu|\xi_{i}^\mu} =1$ and these objects should be understood as equivalence classes defined modulo a phase, ${\cal M} \sim \mathbb{S}^{2d-1}/U(1)$. For $d=2$ we are back to the original sphere, $\mathbb{C}P^1 = \mathbb{S}^2$ - the Bloch sphere. However, this is a special ``exceptional'' case in this family of manifolds and for $d>2$, $\mathbb{C}P^{d-1}$'s are not spheres but have a richer geometry and topology. As we shall see below, it gives rise to a qualitatively different recall protocol in the $SU(d)$ Hopfield networks and their ``exploding'' capacity.

The $SU(d)$ group also provides a nice way to embed ${\cal M}$ into a higher-dimensional sphere and set up a framework to deal with non-linear constraints through a linear Lie-algebraic structure. Consider a neuron $\ket{s_i} \in {\cal M}=\mathbb{C}P^{d-1}$. Define a $D=(d^2-1)$-dimensional real vector (``Bloch magnetization'') as follows
\begin{equation}
\label{vecS}
{\bm s}_i = \braket{s_i|\hat{\bm \lambda}|s_i},
\end{equation}
where $\hat{\bm \lambda}$ is a $D$-dimensional vector of Gell-Mann matrices. A similar definition is for the memories ${\bm \xi}_i^\mu = \braket{\xi_i^\mu|\hat{\bm \lambda}|\xi_i^\mu}$. The algebra of these matrices satisfies a number of useful canonical relations, including:
\begin{equation}
\label{Casimir}
\sum\limits_{a=1}^{D=d^2-1} \hat{ \lambda}_a^2 = \frac{2 (d^2-1)}{d} \hat{I},
\end{equation}
and 
\begin{equation}
\label{Fiertz}
\hat {\bm \lambda}_{\alpha \beta} \cdot \hat {\bm \lambda}_{\alpha' \beta'} = 2 \delta_{\alpha \beta'} \delta_{\alpha' \beta} - 
\frac{2}{d} \delta_{\alpha \beta} \delta_{\alpha' \beta'},
\end{equation}
where $\hat{I}$ is the $d \times d$ identity matrix. Hence, the spin and memory vectors lie on a $(d^2-2)$-dimensional sphere
$$
{\bm s}_i \in {\cal M}=\mathbb{C}P^{d-1} \subset \mathbb{S}^{d^2-2} \subset \mathbb{R}^{d^2-1}.
$$
But for all $d>2$, the manifold ${\cal M}=\mathbb{C}P^{d-1}$ itself is not a sphere but rather a lower-dimensional $2(d-1)$ surface ``hidden'' in the much higher-dimensional sphere through the non-linear constraints. In what follows, we label neurons and memories using both complex ``braket'' notations, ($\ket{s_i}$ and $\ket{\xi_i^\mu}$)  and real vector notations (${\bf s}_i$ and ${\bm \xi}_i^\mu$ with the corresponding elements ${ s}_{ia}$ and ${ \xi}_{ia}^\mu$ labeled by the indices $a,b \in \{1,2,\ldots,d^2-1\}$). The indices $i$ and $j \in \{1,2,\ldots,N\}$ label the individual neurons and $\mu \in \{1,2,\ldots,P\}$ the memories.

\section{$SU(d)$ Hopfield network}
\label{sec:sud}

Using the construction from the previous section, we can now define an $SU(d)$ Hopfield neural network or equivalently Hopfield network on the symmetric space, $\mathbb{C}P^{d-1}$, which has enough algebraic structure to generalize the conventional Hopfield networks, including defining the recall procedure, capacity, replica analysis, etc. Using Eqs.~(\ref{J}), (\ref{HE}), and (\ref{vecS}), we obtain:
\begin{equation}
\label{Ekets}
{\cal E} = -\frac{1}{2N} \sum\limits_\mu \left( \sum\limits_i \braket{s_i| \hat {\bm \lambda}|s_i} \cdot \braket{\xi_i^\mu| \hat {\bm \lambda}|\xi_i^\mu} \right)^2
\end{equation}
We now use the completeness relations for the $SU(d)$ generators (\ref{Fiertz}) to derive 
\begin{equation}
\label{Ekets2}
{\cal E} = -\frac{2}{N} \sum\limits_\mu \left( \sum\limits_i \left| \braket{s_i|\xi_i^\mu} \right|^2 - \frac{N}{d} \right)^2 \equiv - \frac{N}{2} \sum\limits_\mu O_\mu^2
\end{equation}
Note that according to these relations, the norm of the memory and spin vectors  $\left| {\bm s}_i \right|^2=2(1-1/d)$. Hence, in this convention,  the Mattis overlap is
\begin{equation}
\label{O}
 O_\mu = \frac{2}{N} \sum_i \left( \left| \braket{s_i|\xi_i^\mu} \right|^2 - \frac{1}{d} \right) \in \left[ -\frac{2}{d}, \frac{2(d-1)}{d} \right]
\end{equation}
where the upper limit corresponds to a perfect alignment and the lower limit to {a state orthogonal to the memory at every site (a random unrelated state has $O_\mu \sim O(1/\sqrt{N})$}. These numbers have no particular significance and come from standard conventions involving the Gell-Mann matrices. They can be renormalized for the overlap to lie between zero and one, but we proceed as is. 

Now, suppose the state of the system corresponds to a slightly corrupted memory, $\mu$. The asynchronous descent involves updating neurons one at a time. Consider a neuron $i$ and split the overlap into a part that contains it and one that does not. Then we can identically re-write
\begin{equation}
\label{O'}
O_\mu = \frac{2}{N} \left( \left| \braket{s_i|\xi_i^\mu} \right|^2 -1/d \right) + O_\mu^{{(i)}}
\end{equation}
with 
$$
O_\mu^{(i)} = \frac{2}{N} \sum_{j \ne i} \left( \left| \braket{{s_j|\xi_j^\mu}} \right|^2 - \frac{1}{d} \right).
$$
The point of this trivial rewriting is to notice that the first term in Eq.~(\ref{O'}) is $\sim O(1/N)$ and the second term is $\sim O(1)$ but $i$-independent. The recall process is to modify the neuron $i$ to most efficiently lower the energy that depends on it. Hence in the large-$N$ limit, we can focus only on the ``active'' energy that is $i$-dependent and is not infinitesimally small in $1/N$. Plugging Eq.~(\ref{O'}) in Eq.~(\ref{Ekets2}), dropping terms of order $O(1/N)$ (self-interactions) and  $i$-independent terms, the $i$-active energy can be written as
\begin{equation}
\label{EK}
{\cal E}_{\rm active}[s_i] = - 2 \braket{s_i|\hat{K}_i|s_i},
\end{equation}
where $\hat{K}_i$ is a $d \times d$ Hermitian matrix (memory kernel)
\begin{equation}
\label{K}
\hat{K}_i = \sum\limits_\mu O_\mu^{(i)}\,  \ket{\xi_i^\mu} \bra{\xi_i^\mu}
\end{equation}
This structure is known as Rayleigh quotient and is bounded as follows:
$$
\braket{s_i|\hat{K}_i|s_i} \in \left[ \lambda_{\rm min}^{(i)}, \lambda_{\rm max}^{(i)} \right],
$$
where $\lambda_{\rm min}^{(i)}$ and $\lambda_{\rm max}^{(i)}$ are the minimal and maximal eigenvalues of $\hat{K}_i$, which is an elementary fact easily provable by expanding $\ket{s_i}$ in terms of the eigenvectors of $\hat{K}_i$. Since, our goal is to find the most efficient way to lower the energy by updating neuron $i$, it reduces to maximizing the diagonal matrix element of the memory kernel. Obviously, this implies aligning $\ket{s_i}$ along the corresponding top eigenvector 
\begin{equation}
\label{update}
\ket{s_i} \rightarrow \ket{s_i^{\rm new}}=\ket{\chi_{\rm max}^{(i)}}\!: \hat{K}_i \ket{\chi_{\rm max}^{(i)}} = \lambda_{\rm max}^{(i)} \ket{\chi_{\rm max}^{(i)}}
\end{equation}
This is the update rule for the generalized Hopfield networks. Note that for $SU(2)$ this procedure is identical to vector alignment of ${\bf s}_i$ with the mean-field vector produced by other neurons. So, in this case the above sophisticated construction collapses to the familiar equations (\ref{J}) and (\ref{HE}) we started with, and hence is unnecessary. But for $d\ge 3$, it is a qualitatively different structure where the top eigenvector of the memory kernel is better protected from the random matrix noise.

\begin{figure}
\centering
\includegraphics[width=\linewidth]{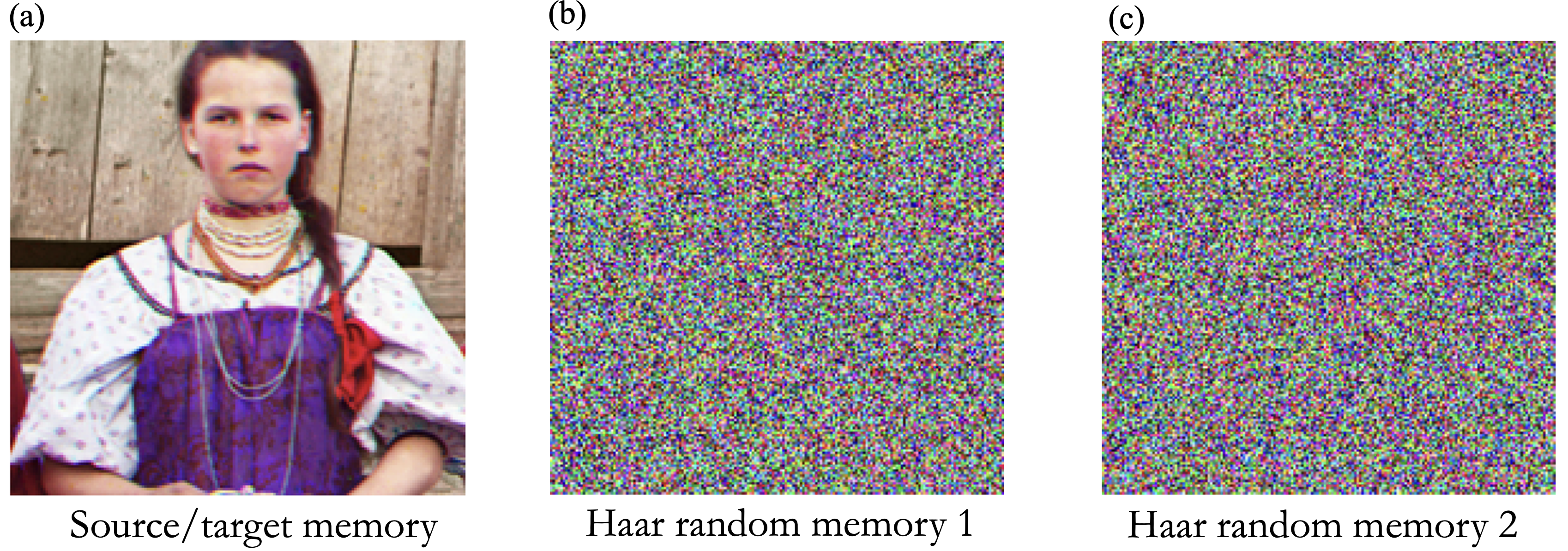}
\caption{Examples of memories stored for the $SU(3)$ image-recall demonstration.
(a) A fragment of a color photo  (we have chosen one of the earliest known color photographs by Prokudin-Gorsky 
taken in 1909 using RGB filters~\cite{ProkudinGorskii}). The image is compressed into a $180 \times 180=32400$-pixel matrix which is flattened into a 
memory vector, ${\bm \xi}^{\mu=0}_i$, where $i$ runs over 32400 components. Each pixel is a neuron which is encoded as a qutrit on
$\mathbb{C}P^2$ through the color-to-qutrit map of Eq.~(\ref{encoding}).
(b),(c) Two Haar-random qutrit memories, shown decoded back to RGB.}
\label{fig:memories}
\end{figure}

\section{Image encoding and recovery with qudit Hopfields}
\label{sec:image}
\begin{figure*}[t]
\centering
\includegraphics[width=\linewidth]{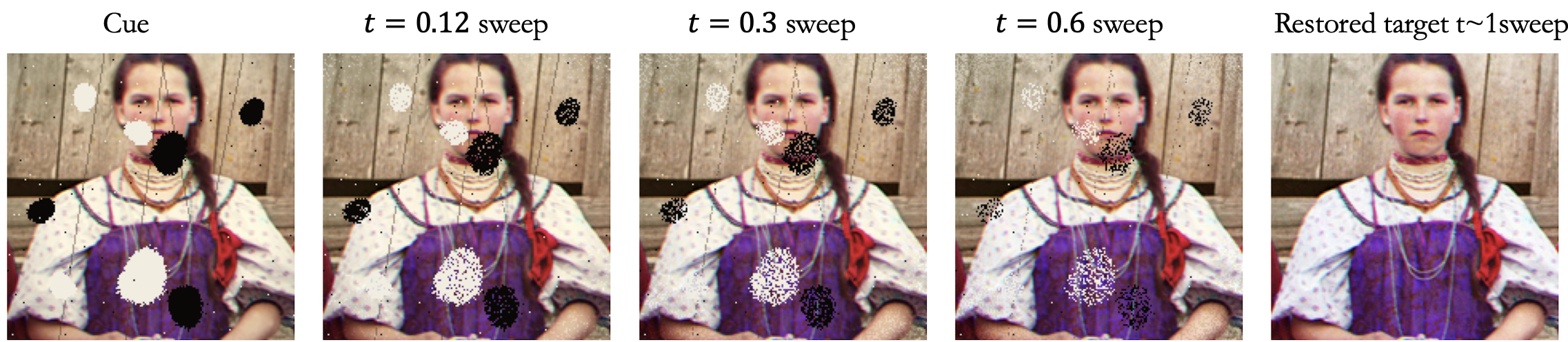}
\caption{Image recall from the corrupted cue using the top eigenvector update
procedure for the $SU(3)$ neural network. The progression illustrates the
``restoration'' of the image as a function of ``time,'' measured in the
fraction of a single sweep ($32400$ individual neuron updates). For this
ultra-low load of the network ($100$ Haar random memories and the target
image) the near perfect restoration is achieved within a single sweep and
then settles to a $100\%$ match.}
\label{fig:recall}
\end{figure*}
To demonstrate memory recall and calculate network capacity of $SU(d)$ Hopfields, we can use any sources of memory neurons. But for the sake of numerical demonstration, we may as well use meaningful images (combined with randomly generated patterns, whenever statistics needs to be collected). We choose the $(R,G,B) \in [0,1]^3$ additive color model for image encoding and start with the simplest non-trivial novel case of $SU(3)$ qudits. This poses an independent interesting challenge of optimal reversible embedding of the three-dimensional ``color cube'' into a $4$-dimensional non-contractible symmetric space, $\mathbb{C}P^2$, with nontrivial topology.  The naive approach of just assigning $\ket{R,G,B}/{\rm Norm}$ to a qutrit is not appropriate because it is not an injection: e.g., white $(1,1,1)$ and grey $(0.5,0.5,0.5)$ colors correspond to the same physical neuron state and black $(0,0,0)$ is not defined at all. {We use the following ad hoc encoder/decoder scheme: (i)~For a pixel $(R,G,B)$, subtract $1/2$ componentwise to center it $\vec{p}=(R-1/2,G-1/2,B-1/2)=(p_r,p_g,p_b)\in[-1/2,1/2]^3$, so that $|\vec p|^2\leq 3/4$. (ii)~Calculate the norm $|\vec{p}|$ and (iii)~define
\begin{equation}
\label{encoding}
(R,G,B) \rightarrow \ket{\xi}=(q_1,q_2,q_3), \mbox{ with }
\end{equation}
\[
q_1=\sqrt{\frac{1+\sqrt{1-\frac{4}{3} |{\bf p}|^2}}{2}}, \,
q_2= \frac{p_r + i p_g}{\sqrt{3}\,q_1}, \,
q_3=\frac{p_b}{\sqrt{3}\,q_1}.
\]
With this normalization, $\braket{\xi|\xi}=1$ for every pixel in the RGB cube.

The inverse, restricted to the encoded RGB submanifold, is the qutrit-to-pixel decoder
\begin{equation}
\begin{pmatrix} R \\ G \\ B \end{pmatrix}
= \frac{1}{2}\begin{pmatrix} 1 \\ 1 \\ 1 \end{pmatrix}
+ \sqrt{3}\,
\begin{pmatrix}
\operatorname{Re}\!\left(\overline{q_1}\,q_2\right) \\[2pt]
\operatorname{Im}\!\left(\overline{q_1}\,q_2\right) \\[2pt]
\operatorname{Re}\!\left(\overline{q_1}\,q_3\right)
\end{pmatrix},
\end{equation}
followed, if necessary, by clipping each component to $[0,1]$. That is, if a component is below $0$, set it to $0$; if a component is above $1$, set it to $1$.} There are probably more convenient encoder/decoder schemes, but for our purposes this prescription works. {It satisfies minimal conditions for the encoded images: the inverse is exact on the encoded RGB submanifold and the decoder is independent of the overall phase of the qutrit (``gauge-invariant'').}

To verify that the encoding, decoding, and recall procedures work as intended we first test them on a $SU(3)$ Hopfield network with minimal capacity. To illustrate the recall, we choose the image in Figure~{\ref{fig:memories}}(a) [a compressed fragment of an iconic image from 1909]. We then flatten the image into a vector and convert it into a qutrit or equivalently a point on the $\mathbb{C}P^2$ manifold as described above. We also generate $100$ Haar random memories; see examples in Figures~\ref{fig:memories}(b) and (c). We then corrupt the source image as shown in Fig.~\ref{fig:recall} and run the $SU(3)$ recall protocol on the network with a few intermediate network states shown in the Figure. We observe a very fast recovery to the nearly perfect target image within a single sweep (i.e., going over all $32400$ neurons in a random order) with minimal cross-talk. 

We next examine the network capacity for $SU(2)$, $SU(3)$, and $SU(4)$ Hopfields. This is a more computationally expensive exercise than low-load single-image recovery. We use the eigenvector asynchronous recall rule derived above with random memories.  For each system size $N$ and load $\alpha=P/N$,  $P$ independent Haar-random memories on the relevant manifold are generated.   retrieval overlap is defined as 
$$
\bar{m}_\mu = \frac{d}{2(d-1)}\, O_\mu,
$$
so that $\bar{m}_\mu=1$ for perfect recall and $\bar{m}_\mu\simeq 0$ for an unrelated state.  For each value of $\alpha$, we average the final retrieval {overlap} over independent disorder realizations.  For computational estimate of the critical capacity, $\alpha_c$, we need to make a subjective choice of  the critical overlap threshold, which we set to $\bar{m}_{\rm th}=1/2$ ($\alpha_c$ is well-defined in the thermodynamic limit, where it corresponds to a genuine transition).  The computational estimate is obtained from a linear extrapolation of the critical $\alpha_c(N)$ versus $1/N$.  This procedure gives $\alpha_c\simeq0.05$ for $SU(2)$, consistent with the earlier estimates for the Heisenberg/vector Hopfield model~\cite{Nicoletti2025,Barney2026}. This is much lower than the capacity of the canonical binary network. For $d \ge 3$, we observe a strong order-of-magnitude enhancement in the critical capacity for $SU(d)$ Hopfields:  $\alpha_c \approx 0.62$ for $SU(3)$ and $\alpha_c \approx 2.41$ for $SU(4)$. 

In addition to the increased capacity of the entire network, each $SU(d)$ neuron is a rich object for information encoding. Even $SU(3)$ neurons are described by $4$ independent real parameters (albeit subject to non-linear geometric constraints to lie on the manifold, $\mathbb{C}P^2$). For example, the ad hoc $(R,G,B)$ image encoding we used above leaves another phase-like degree of freedom unused. One way to make it useful is to add a ``clock'' to the image such that a single qutrit could describe a periodic process rather than a static picture.

\section{The nature of the enhanced critical capacity}
\label{sec:capacity}

\subsection{Qualitative picture: Spiked memory kernel with random matrix noise}
 Before proceeding with the more rigorous derivation, we provide qualitative arguments for the enhanced stability of top eigenvector recall against fluctuations, which is associated with a gap protecting it from tilt (at least at small loads). Consider a state of the $SU(d)$ network close to memory $\ket{\xi^1_i}$, or equivalently $O_{\mu =1} \sim O(1)$. The memory kernel matrix (\ref{K}) can be written as:
\begin{equation}
\label{KBBP}
\hat{K}_i = O_1^{(i)} \ket{\xi_i^1} \bra{\xi_i^1} + \sum_{\mu \ge 2} O_\mu^{(i)} \ket{\xi_i^\mu} \bra{\xi_i^\mu}
\end{equation}
Note that the first term is a large ``spiked'' matrix and the sum includes $\alpha N \gg 1$ small (i.e., $|O_\mu^{(i)}| \ll 1$ for all $\mu \ge 2$) random matrices. Per the central limit theorem, we expect the latter to be close to one drawn from the Gaussian unitary ensemble~\cite{Mehta2004}. Qualitatively, we can write
\begin{equation}
\label{KBBP2}
\hat{K} \approx m \ket{1} \bra{1} + {\sqrt{\alpha}}\, C(d) \hat{G},
\end{equation}
where $m\sim 1$, $\hat{G}$ is a GUE random matrix, $\alpha =P/N$ is the load, and $C(d)$ is a constant (expected to be order one for $d\sim 3$  and to decay as $1/d^2$ for $d \gg 1$). If $\alpha$ is small, there is a gap protecting the top eigenvector of the memory kernel. We expect a transition where the random matrix band and the eigenvalue of the signal overlap, similar to the Baik-Ben Arous-P{\'e}ch{\'e} (BBP) transition~\cite{BBP2005} in random matrix theory (RMT). One can rigorize this construction using RMT and elements of cavity methods (or equivalently  the Thouless-Anderson-Palmer approach that takes into account the Onsager reaction term) and estimate the critical capacity this way, but we proceed using the replica technique below. However, equation~(\ref{KBBP2}) is useful to gain intuition about the origin of the enhanced stability of the $SU(d)$ networks and their increasing capacity with increasing the manifold dimension. This is in sharp contrast to the drop of the critical capacity in vector Hopfield networks on spheres, $\mathbb{S}^n$ with $\alpha_c \propto 1/n$~\cite{Nicoletti2025,Bolle2003spherical}, which is related to the ``gapless'' nature of the vector alignment.

\begin{figure}[t]
\centering
\includegraphics[width=0.8\linewidth]{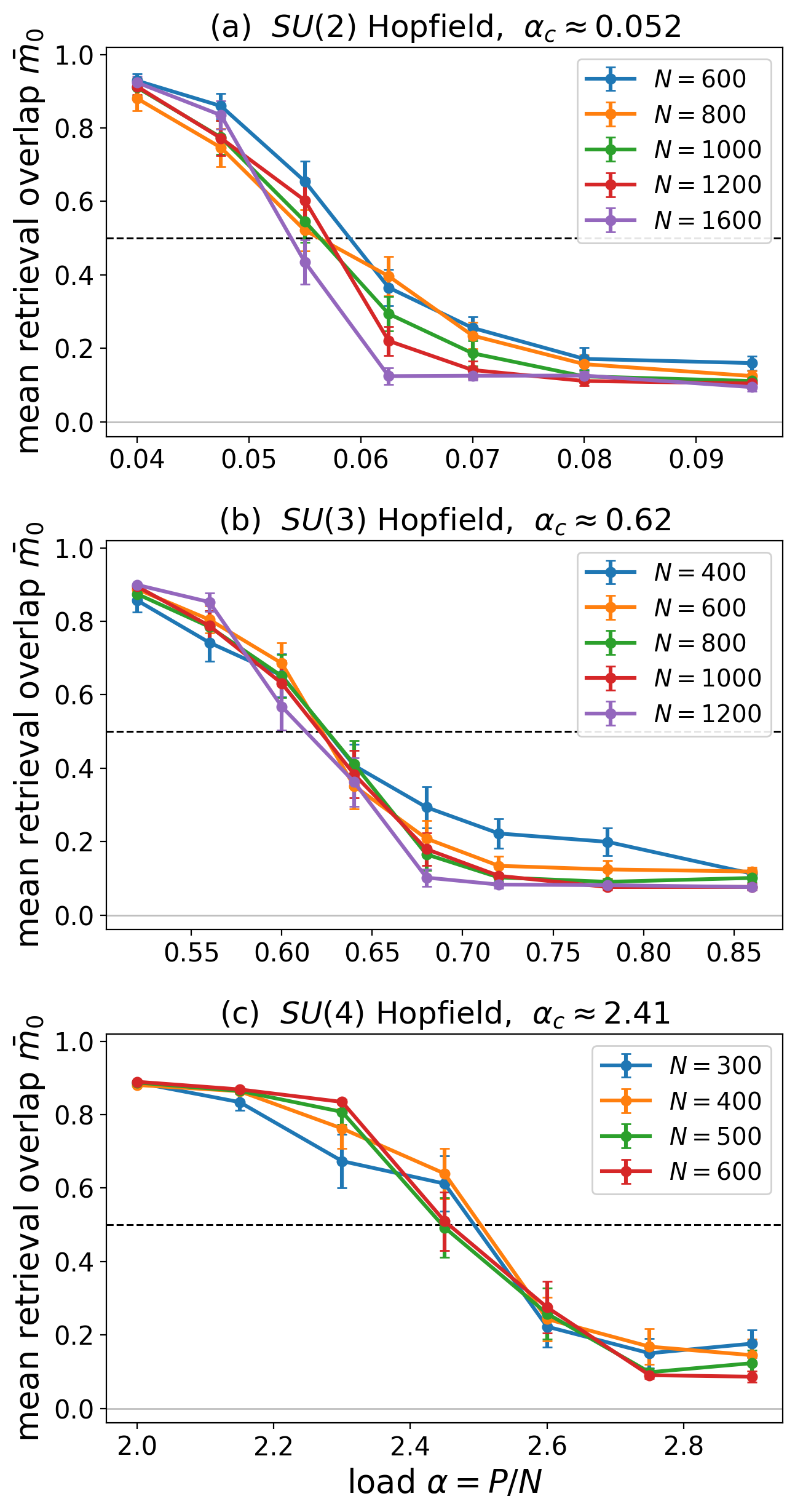}
\caption{Numerical {simulations} of the $SU(d)$ recall  for $d=2$, $3$, and $4$ for {$P$} Haar random memories on $\mathbb{C}P^{d-1}$. Mean {retrieval overlap $\bar{m}_0 = \frac{d}{2(d-1)}\, O_0$ with $O_{\mu}$ defined in Eq.~(\ref{O}) is plotted as a function of the network load. The dashed lines mark the threshold $\bar{m}_{\rm th}=1/2$}. We observe a sharp deterioration of the image recovery process and estimate the critical capacity, $\alpha_c$ from this behavior. It shows an order of magnitude enhancement starting with $d=3$.}
\label{fig:capacity}
\end{figure}

\subsection{Replica analysis}
{This section maps $SU(d)$ Hopfield networks to a statistical-mechanical problem and estimates the critical capacity using the replica and Hubbard-Stratonovich method~\cite{EA1975,SK1975}. As before, define the following $D=(d^2-1)$-dimensional vectors ${\bf s}_i$ and ${\bm \xi}^\mu_i \in \mathbb{R}^{D}$. The former are statistical mechanical degrees of freedom and the latter  (memories) are quenched disorder; both lie on the symmetric space $\mathbb{C}P^{d-1}\subset\mathbb{R}^{D}$. The partition function corresponding to the energy (\ref{Ekets2}) is
\begin{equation}
\label{Z1}
Z = \int \prod\limits_{i=1}^{{N}} d\mu(s_i)  \exp{\left[ \frac{\beta N}{2} \sum\limits_{\mu =1}^P O_\mu^2 \right]},
\end{equation}
where $d\mu(s_i)$ is a measure on $\mathbb{C}P^{d-1}$, $O_\mu = N^{-1} \sum\limits_i {\bf s}_i \cdot {\bm \xi}_i^\mu$, and $T = 1/\beta$ is the temperature (which will be taken to zero in the end).}

\begin{figure*}[t]
\centering
\includegraphics[width=0.9\linewidth]{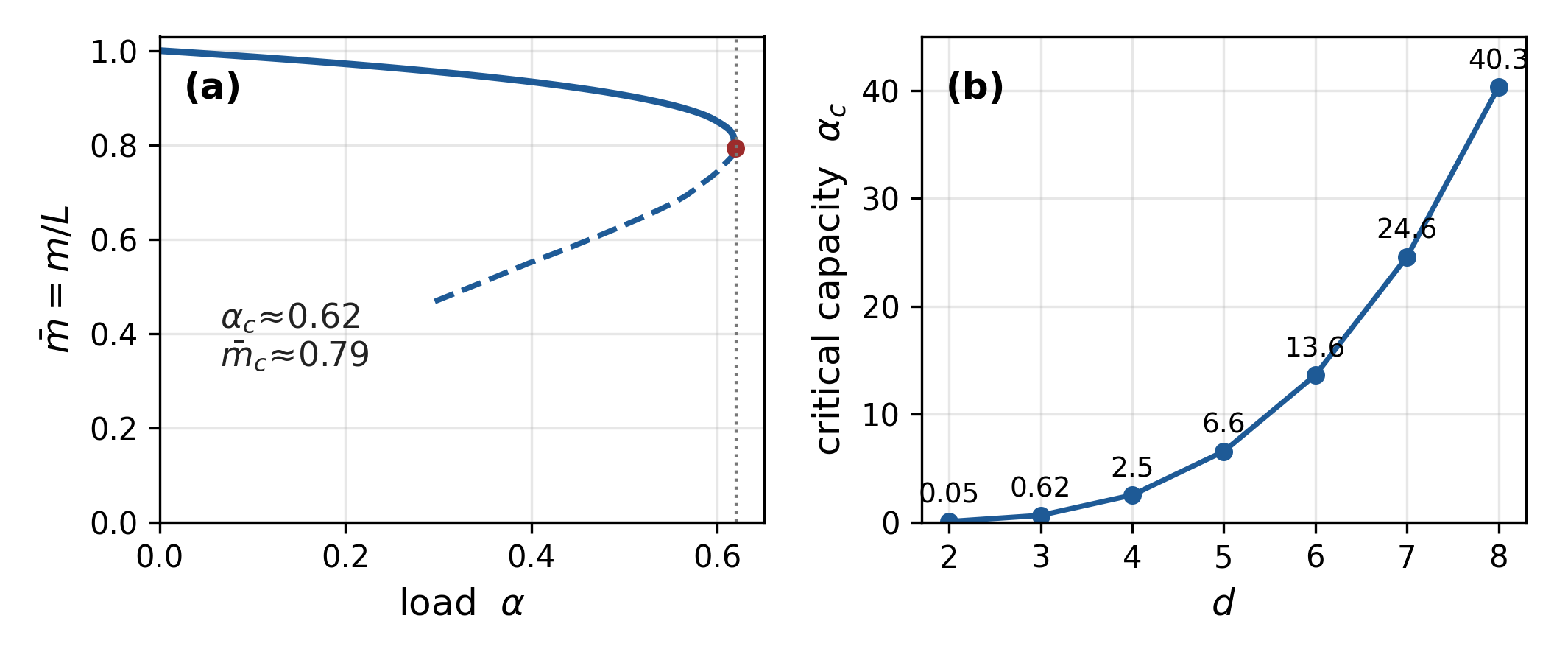}
\caption{(a)~Replica-symmetric retrieval overlap $\bar m=m/L$ versus load $\alpha$ for the $d=3$
($\mathbb{C}P^2$, qutrit) network, from the $T=0$ self-consistent equations~(\ref{T0}) averaged over
$\sim\!10^4$ samples of the matrix field $\hat K$. The stable (solid) and unstable (dashed) branches merge
at the critical load that defines $\alpha_c\simeq0.62$. (b)~Critical capacity $\alpha_c$ as a function of the qudit dimension $d$, obtained from  the $T=0$ replica-symmetric equations~(\ref{T0}). The values $\alpha_c\simeq0.05,0.62,2.5,6.6,13.6,24.6,40.3$
for $d=2,\dots,8$ rise steeply, exceeding unity for $d\ge4$ and far above the binary value $0.138$.}
\label{fig:malpha}
\end{figure*}

{The quenched free energy $f=-\lim_{N\to\infty}(\beta N)^{-1}\,\overline{\ln Z}$ follows from the replica
identity $\overline{\ln Z}=\lim_{n\to0}n^{-1}(\overline{Z^n}-1)$, where $\overline{Z^n}$ is the disorder
average of $n$ copies (replicas $a=1,\dots,n$) and the memories are Haar-distributed on $\mathbb{C}P^{d-1}$,
\begin{equation}
\label{Haar}
\Ev[\xi^\mu_{ia}]=0,\quad \Ev[\xi^\mu_{ia}\xi^\nu_{jb}]=C\,\delta^{\mu\nu}\delta_{ij}\delta_{ab},\quad
C=\frac{2}{d(d+1)},
\end{equation}
with $|{\bm\xi}^\mu_i|^2=L\equiv2(d-1)/d$ [Eq.~(\ref{Fiertz})]. Each square in $\overline{Z^n}$ is linearized
by a Hubbard-Stratonovich (HS) transformation that introduces the overlap order parameter $m^a_\mu$,
\begin{equation}
\label{HS1}
e^{\frac{\beta N}{2}(O^a_\mu)^2}=\sqrt{\tfrac{\beta N}{2\pi}}\!\int\!dm^a_\mu\,
e^{-\frac{\beta N}{2}(m^a_\mu)^2+\beta m^a_\mu\sum_i{\bm s}^a_i\cdot{\bm\xi}^\mu_i}.
\end{equation}
Retrieval of memory $1$ corresponds to a condensed overlap $m^a_1\equiv m^a=O(1)$, the remaining $m^a_{\mu\ge2}$
being of order $N^{-1/2}$. Averaging the $\alpha N$ uncondensed memories to second order with
Eq.~(\ref{Haar}) gives
\begin{equation}
\label{xtalk}
\overline{\exp\!\Big[\beta\!\sum_a m^a_\mu\!\sum_i{\bm s}^a_i\!\cdot\!{\bm\xi}^\mu_i\Big]}
=\exp\!\Big[\tfrac{\beta^2 C N}{2}\!\sum_{ab}m^a_\mu m^b_\mu\,q^{ab}\Big],
\end{equation}
where the replica overlap is
\begin{equation}
\label{qab}
q^{ab}=\frac1N\sum_i{\bm s}^a_i\cdot{\bm s}^b_i,\qquad q^{aa}=L .
\end{equation}
Integrating out the Gaussian $m^a_{\mu\ge2}$ produces the factor
$\exp[-\tfrac{\alpha N}{2}\Tr\ln(\delta^{ab}-\beta C q^{ab})]$. We then promote $q^{ab}$ to an integration
variable with conjugate $r^{ab}$, and use the $SU(d)$ freedom to align the condensed memory with a
fixed reference ${\bm v}_0$ (the Bloch vector of $\ket{\psi_0}=(1,0,\dots,0)^T$, with $|{\bm v}_0|^2=L$).
This brings the replicated partition function to the following form $\overline{Z^n}=\int dm\,dq\,dr\,e^{NS_n}$ with
\begin{align}
\label{action}
S_n={}&-\frac\beta2\sum_a(m^a)^2-\frac\alpha2\Tr\ln(\delta^{ab}-\beta C q^{ab})\nonumber\\
&{}-\beta\sum_{a<b}r^{ab}q^{ab}+\ln\zeta_n,\nonumber\\
\zeta_n={}&\int\prod_a d\mu(s^a)\,
\exp\!\Big[\beta\!\sum_a m^a\,{\bm v}_0\!\cdot\!{\bm s}^a+\beta\!\!\sum_{a<b}\!\!r^{ab}\,{\bm s}^a\!\cdot\!{\bm s}^b\Big].
\end{align}

We confine ourselves to the replica-symmetric ansatz/approximation as follows $m^a=m$, $q^{a\ne b}=q$, $r^{a\ne b}=r$. A second HS transformation,
now with a single $D$-component field ${\bm z}$ shared by all replicas, linearizes the inter-replica
coupling $\sum_{a<b}{\bm s}^a\!\cdot\!{\bm s}^b$,
\begin{equation}
\label{HS2}
e^{\frac{\beta r}{2}(\sum_a{\bm s}^a)^2}=\int\!{\cal D}{\bm z}\,
e^{\sqrt{\beta r}\,{\bm z}\cdot\sum_a{\bm s}^a},
\end{equation}
where we introduced the Gaussian measure ${\cal D}{\bm z}=\prod_{c=1}^{D}(dz_c/\sqrt{2\pi})\,e^{-z_c^2/2}$ for brevity. This
decouples the sites/neurons, $\zeta_n=\int{\cal D}{\bm z}\,\zeta_1^{\,n}$, leaving each neuron as a single qudit in
an effective kernel $\hat K$,
\begin{equation}
\label{zeta1}
\zeta_1=\int d\mu(s)\,e^{\beta\braket{s|\hat K|s}},\quad
\langle{\cal O}\rangle=\frac1{\zeta_1}\int d\mu(s)\,{\cal O}\,e^{\beta\braket{s|\hat K|s}},
\end{equation}
with $\langle\cdot\rangle$ the single-site Gibbs average. The kernel is the $d\times d$ Hermitian matrix
\begin{align}
\label{Hfield}
\hat K&=m({\bm v}_0\!\cdot\!\hat{\bm\lambda})+\sigma({\bm z}\!\cdot\!\hat{\bm\lambda}),\nonumber\\
{\bm z}&\sim{\cal N}(0,\mathbb{1}_{D}),\qquad \sigma^2=r/\beta .
\end{align}
By Eq.~(\ref{Fiertz}) the signal term is a ``spiked matrix,''
${\bm v}_0\!\cdot\!\hat{\bm\lambda}=2\ket{\psi_0}\bra{\psi_0}-\tfrac2d\hat I$, while
$\sigma({\bm z}\!\cdot\!\hat{\bm\lambda})$ is a GUE matrix; thus $\hat K$ is the mean-field
(disorder-averaged) form of the instance kernel $\hat K_i$, cf.\ Eq.~(\ref{KBBP2}). Extremizing $S_n$ in the
$n\to0$ limit gives the saddle point
\begin{align}
\label{saddle}
m&=\Ev_{\bm z}\big[{\bm v}_0\!\cdot\!\langle{\bm s}\rangle\big],\qquad
q=\Ev_{\bm z}\big[\langle{\bm s}\rangle\!\cdot\!\langle{\bm s}\rangle\big],\nonumber\\
\sigma^2&=\frac{\alpha\,C^2 q}{[\,1-\beta C(L-q)\,]^2}.
\end{align}
The structure is identical to the binary Hopfield model~\cite{AGS1985}, the manifold entering through the parameters
$C$ and $L$. Note that the factor $[1-\beta C(L-q)]^{-2}$ is the Onsager reaction~\cite{TAP1977}.

In the recall limit $T\to0$, the Boltzmann weight $e^{\beta\braket{s|\hat K|s}}$ is determined entirely by the
top eigenvector of $\hat K$. With the spectral decomposition
\begin{equation}
\label{spec}
\hat K\ket{k}=e_k\ket{k},\qquad e_1>e_2\ge\dots\ge e_d
\end{equation}
($k=1,\dots,d$), the magnetization aligns with the top eigenvector, $\ket{\rm top}$,
$\langle{\bm s}\rangle\to{\bm s}_{\rm top}$. The combination $\beta(L-q)$ remains finite and tends to
$\Ev_{\bm z}[\chi]$, where the single-site susceptibility $\chi=\sum_a\partial\langle s^a\rangle/\partial h_a$
follows in closed form from first-order perturbation theory and the completeness relation~(\ref{Fiertz}),
\begin{equation}
\label{chi}
\chi=4\sum_{k\ne1}\frac{1}{e_1-e_k}\ge0 .
\end{equation}
The closed replica-symmetric equations at $T=0$ are then
\begin{align}
\label{T0}
m&=\Ev_{\bm z}\!\Big[2\,|\!\braket{\psi_0|{\rm top}}\!|^2-\tfrac2d\Big],\nonumber\\
\sigma^2&=\frac{\alpha\,C^2 L}{(1-g)^2},\qquad g=C\,\Ev_{\bm z}[\chi].
\end{align}

We solve these by diagonalizing $\hat K$ over $\sim\!10^4$ samples of ${\bm z}$ and iterating $(m,g)$ to a
fixed point. The retrieval overlap $\bar m=m/L$ measures the quality of the recovered state ($\bar m=1$ for
perfect recall, $\bar m\to0$ for an unrelated state). Below $\alpha_c$ the equations admit two retrieval
solutions, shown in Fig.~\ref{fig:malpha}(a): a {stable} branch at high $\bar m$, which is a free-energy
minimum and the state actually reached by the recall dynamics, and an {unstable} branch at lower
$\bar m$, which is the boundary of that minimum's basin of attraction, i.e.\ the smallest cue overlap from
which the network still flows towards the memory. As $\alpha$ increases the two branches approach and annihilate. 
This bifurcation defines $\alpha_c$, and the overlap there - the critical magnetization $\bar m_c$ - 
remains finite ($\bar m_c\approx0.8$ at $d=3$). In the thermodynamic limit, the retrieval experiences as 
a discontinuous jump rather than fading to zero. We observe a strong enhancement in critical capacity with increasing $d$
as shown in Fig.~\ref{fig:malpha}(b). 

For $d=2$ the kernel reduces to $\hat K={\bm h}\!\cdot\!\hat{\bm\sigma}$, recovering vector alignment, and
Eq.~(\ref{T0}) returns the low Heisenberg value $\alpha_c\simeq0.05$. For $d=3$, we obtain $\alpha_c\approx0.62$,
in agreement with the direct simulation. Solving Eqs.~(\ref{T0}) for $d=2,\dots,8$ 
shows - Fig.~\ref{fig:malpha}(b) - $\alpha_c$ growing steadily with $d$, crossing unity already at $d=4$ and reaching $\alpha_c\approx40$ at $d=8$, opposite to the $\alpha_c\propto1/n$ decay of the vector networks on the spheres $\mathbb{S}^n$. We note that the large-$d$ numbers may acquire corrections due to replica symmetry breaking~\cite{Parisi1979}, which we have not explored here.}

\begin{figure*}[t]
\centering
\includegraphics[width=0.8\linewidth]{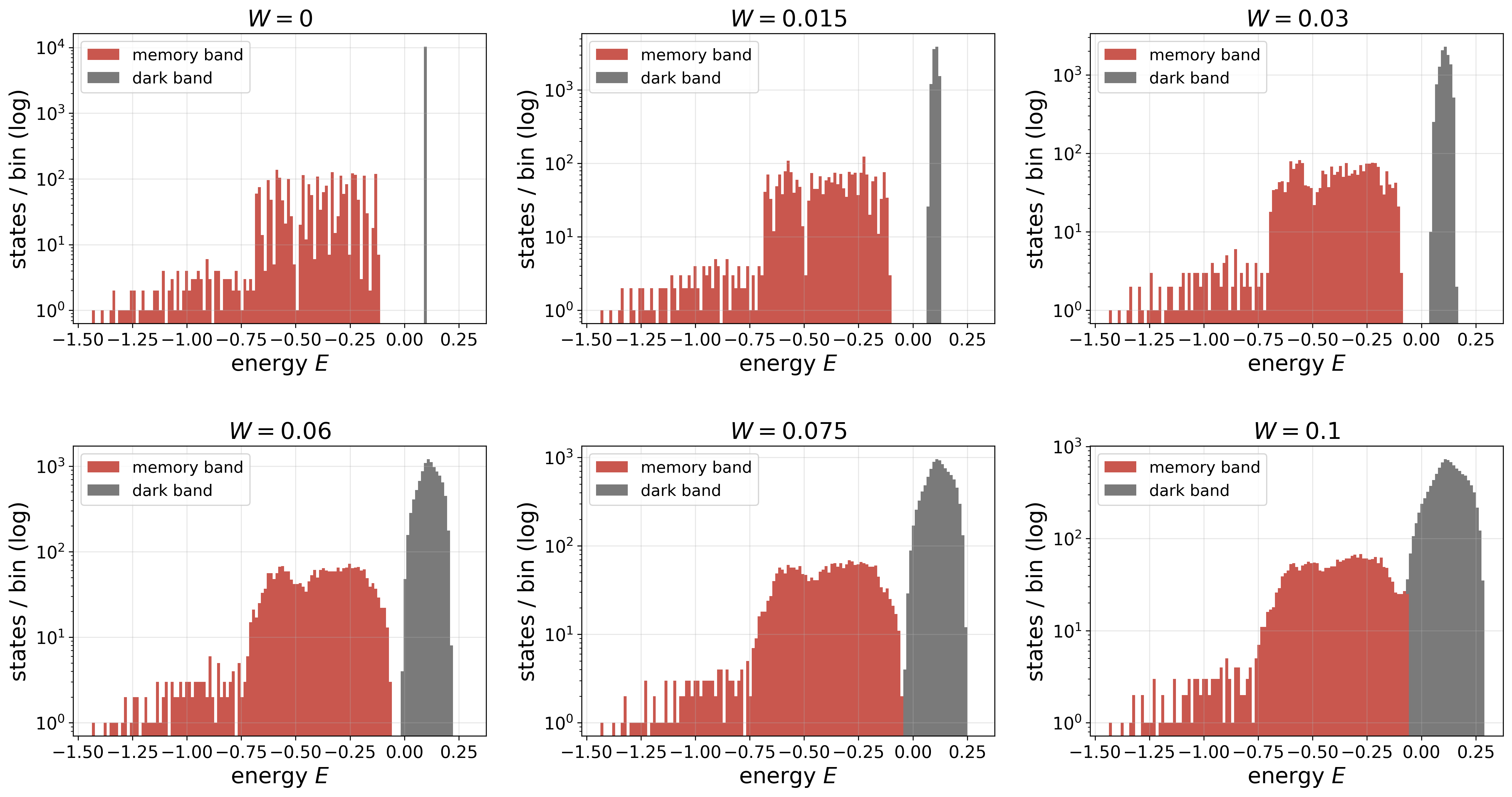}
\caption{Density of states of the quantum XY Hopfield model in the two-magnon sector as a function of the transverse field. The parameters are $N=160$ and $P=16$, corresponding to a load of $10\%$. The magnetic field is taken to be $h_i = W u_i$, where the random variables $u_i$ are drawn uniformly from the interval $[-1,1]$. In the absence of a magnetic field, a flat (dark) band is observed, originating from the zero modes of the Hebb matrix. As the field strength increases, the dark band broadens and eventually merges with the memory band.}
\label{fig:DOS2}
\end{figure*}

\section{Generalized quantum Hopfields}
\label{sec:quantum}

Having established the underlying mathematical structure of classical Hopfield networks on symmetric spaces and demonstrated their potential for both interesting novel mathematical structures and applications, we briefly comment on the possible platforms for their simulation in physical systems. Ironically, the most promising class of system for this are quantum systems, which natively contain $SU(d)$ degrees of freedom. Examples of those are multi-level ultracold alkaline-earth atoms~\cite{Gorshkov2010,Zhang2014}, hyperfine structures in trapped-ion systems~\cite{Ringbauer2022,Senko2015}, and multi-component cold-atom systems embedded in cavities~\cite{Marsh2025,Marsh2021}. Especially the latter have already demonstrated promise in  experimentally realizing both spin glass structures and Hopfield-type associative memory~\cite{Gopalakrishnan2011,Gopalakrishnan2012}. Another possible platform is an interconnected network of small few-qubit quantum computers - each realizing a qudit. In all these candidate systems, we need to delineate truly many body physics, which may involve many-qudit entanglement and active error correction, from emergent quasiclassical mean-field dynamics that as we shall see below will bring these systems back to the classical description on symmetric spaces.  

Given our Lie-algebraic construction, it is straightforward to quantize the generalized Hopfield networks~\cite{RotondoOpen2018,Rebentrost2018,FiorelliPottsHopfield2022,Bodeker2023}. For example, consider the classical energy functional  
\begin{equation}
\label{Ecl}
{\cal H}_{\rm cl} = - \frac{1}{2} \sum\limits_{i \ne j} J_{ij} \braket{s_i|\hat{\bm \lambda}|s_i} \cdot \braket{s_j|\hat{\bm \lambda}|s_j},
\end{equation}
where $\hat{\lambda}_a$ are $d^2-1$  generators of $SU(d)$, which are Hermitian $d \times d$ matrices satisfying the canonical commutation relations~(\ref{cr}) and the $\ket{s} \in {\cal M} =SU(d)/U(d-1)$ are the Perelomov coherent states - the classical vectors on the symmetric space where the classical Hopfields are defined. The natural quantization of the model is to simply replace the classical vectors on neurons, $i,j$, with the corresponding operators 
\begin{equation}
\label{Hq}
\hat{\cal H} = - \frac{1}{2} \sum\limits_{i \ne j} J_{ij}\, \hat{\bm \lambda}_i \cdot \hat{\bm \lambda}_j
\end{equation}
Note that the fully random version (where $J_{ij}$ are iid Gaussian couplings) is the original Sachdev-Ye model of quantum spin glass~\cite{SachdevYe1993}, which is solvable in the large-$d$ limit (this model is not to be confused with the Sachdev-Ye-Kitaev model of randomly coupled fermions~\cite{Rosenhaus2019} which is fully chaotic). 

\begin{figure*}[t]
\centering
\includegraphics[width=0.8\linewidth]{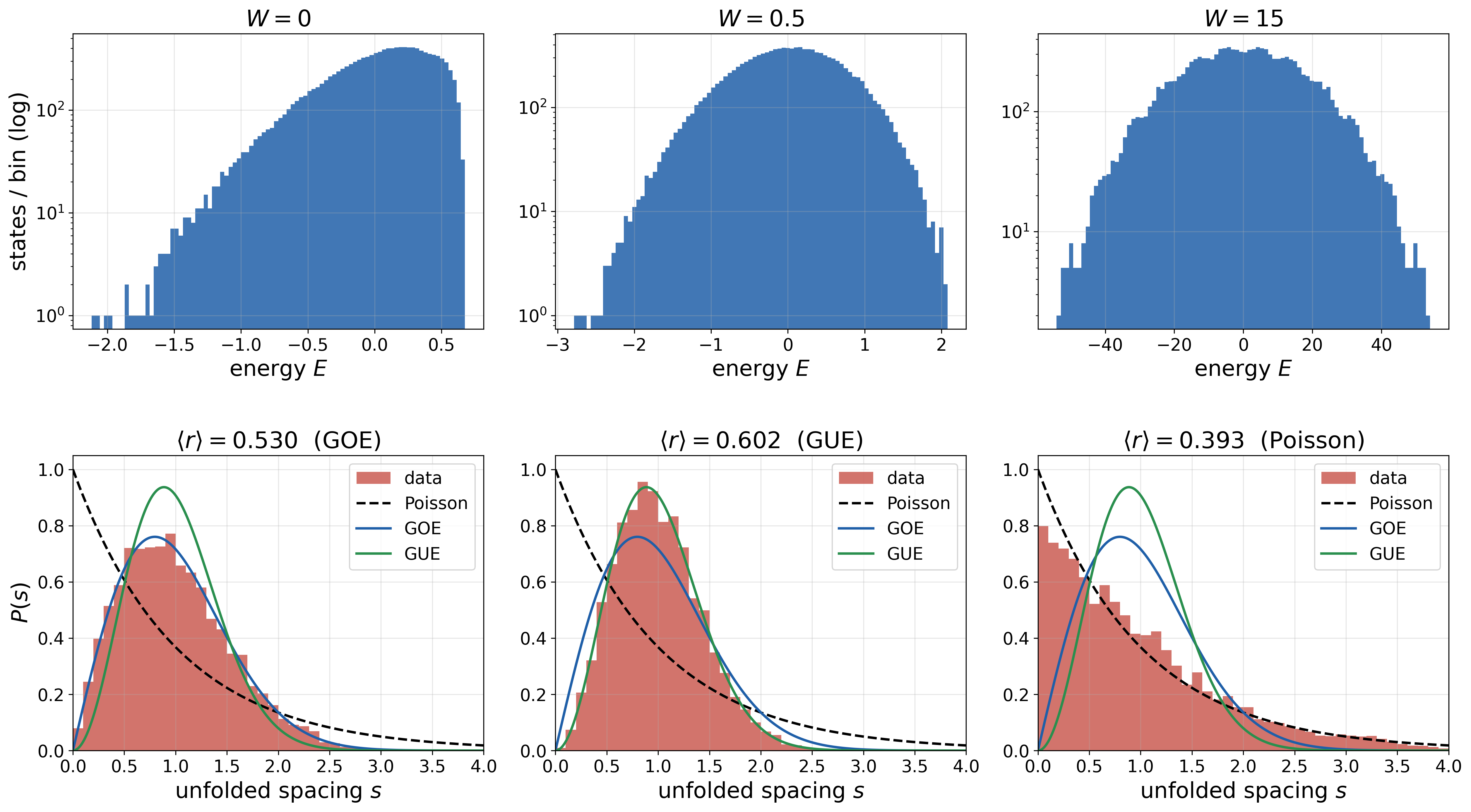}
\caption{Analysis  of the interacting many-body spectrum of the quantum $XY$ Hopfield model at half-filling and as a function of the transverse field. The parameters are $N=16$ and $P=3$ with the Hilbert space size, $n_{\rm Hilbert} = 12870$. Density of states exhibits a crossover to a Gaussian. The  level statistics starts GOE quantum chaotic  and evolves into GUE with the field. At very large field,  a many-body-localization-like crossover is observed, which we believe to be a finite-size artifact.}
\label{fig:DOS12}
\end{figure*}

There are several approaches to studying this class of model (\ref{Hq}). If the primary interest is its phase diagram, all-to-all couplings (either random Gaussian or structured memories) make dynamical mean-field theory a reliable tool for any $d$. As mentioned, large-$d$ models are analytically solvable using $1/d$-expansion. The spin-glass physics is also accessible through the quantum generalization of the Thouless-Anderson-Palmer or replica techniques, where the relevant order parameters and Hubbard-Stratonovich terms are promoted to imaginary-time-dependent fields. 

This paper focuses on the Hopfield-like structures, where the coupling is a structured Hebbian memory matrix. In the context of its quantum formulation, we will be interested in the structure of the many-body spectra and the possible physical (rather than algorithmic) memory recall mechanisms. The appropriate tools here are exact diagonalization techniques and quasiclassical dynamics with dissipation. This research scope puts stringent constraint on the models that can be handled via exact computations: larger-$d$ SU(d) models are practically inaccessible. Hence, we shall study exactly finite-size minimal quantum models and will extrapolate select results to the generic $SU(d)$ case. 

\section{Quantum $XY$-Hopfield network}
\label{sec:xy}

We consider the simplest quantum Hopfield model on a circle in a transverse field (which we may set to zero if we so wish) formulated in terms of the standard spin-$1/2$ matrices $\hat{\bf S} = \hat{\bm \sigma}/2$:
\begin{equation}
\label{QXY}
\hat{\cal H}_{XY} = -\frac{1}{2} \sum\limits_{i \ne j} J_{ij} \hat{S}_i^+ \hat{S}_j^- + \sum\limits_i h_i \hat{S}_i^z,
\end{equation}
with memories defined as phase-like degrees of freedom
\begin{equation}
\label{MXY}
J_{ij} = \frac{1}{N} \sum\limits_{\mu=1}^P {\omega}_i^\mu \overline{\omega_j^\mu} \equiv 
\frac{1}{N} \sum\limits_{\mu=1}^P e^{i(\theta_i^{\mu} - \theta_j^{\mu})}
\end{equation}

Interestingly, this model is equivalent to the generalized Richardson/pairing model of superconductivity considered by the author in Refs.~\cite{Galitski2010pairing,SedrakyanGalitski2010}. For the single-memory $P=1$ case, it becomes the standard single-channel reduced BCS model of superconductivity, which is integrable~\cite{Richardson1963,Dukelsky2004} and where the mean-field approach is exact in the thermodynamic limit (and it becomes reliable at pretty modest particle numbers). Due to all-to-all interactions, we expect the mean-field approach to remain applicable for the non-integrable multi-memory case as well (which in the superconductivity language would correspond to multiple pairing channels). This mean field in effect makes the model classical again and it provides a path to constructing physical real-time recall dynamics for quasiclassical phasor degrees of freedom  that could be generalized to other more complex networks. 

But first we examine the many-body spectral properties of the quantum Hamiltonian (\ref{QXY}). Notice that $[\hat{\cal H}_{XY}, \sum_i\hat{S}_i^z] = 0$ and hence the different $k$-magnon sectors are decoupled. The one magnon sector is in effect single-particle and it largely replicates the spectrum of the Hebb matrix (\ref{MXY}). The regime we study, $P<N$, leads to a finite rank of the matrix and hence  $(N-P)$ zero modes (for zero transverse field) - a ``flat band.'' What is interesting is that ``mesoscopic'' few-magnon sectors retain the ``flat band'' in the many-body spectrum. 

Figure~\ref{fig:DOS2} shows the many-body density of states obtained via exact diagonalization for the two-magnon sector for $N=160$ and $P=16$ and sweeping the magnetic field scale $W$ as labeled ($h_i = W u_i$ where $u_i$ is uniformly distributed on $[-1,1]$). 
At zero field, we observe a  ``memory band''  and the ``dark''/flat band positioned exactly at $E_{\rm flat} = P/N= \alpha$. This is because Hamiltonian (\ref{QXY}) is rewritten, up to the constant $-\frac{1}{2}\sum_i h_i$, as
\begin{equation}
\label{QXY2}
\hat{\cal H}_{XY} = -\frac{1}{2N} \sum\limits_{\mu = 1}^P \hat{B}_\mu^\dagger \hat{B}_\mu + \sum\limits_i \left( \frac{\alpha}{2} +h_i \right) \hat{k}_i,
\end{equation}
where $\hat{B}_\mu = \sum\limits_i \bar{\omega}_i^\mu \hat{S}^-_i$ and $\hat{k}_i=S^z_i + 1/2$. At $h_i=0$, the last sum in (\ref{QXY2}) is just $k \alpha/2$, where $k$ is the magnon number. I.e.,  $E_{\rm dark}=0.1$ in our case. The dark states are those annihilated by all patterns:
$$
\hat{B}_\mu \ket{\rm dark} = 0, \forall \mu
$$
For $N=160$, the total size of the two-magnon Hilbert space is $n_{\rm Hilbert}={N \choose 2} = 12720$, and we find that a large number of them  $n_{\rm dark} = 10280$ are dark. The rest $n_{\rm mem} = 2440$ form the memory band (red region in Figure~\ref{fig:DOS2}), which we find to exhibit strong level repulsion (not shown). As we increase the field, the dark band broadens (grey region in Figure~\ref{fig:DOS2}) and eventually merges with the memory band around $W \sim 0.1$. The dark band eventually absorbs the memory band entirely and shows a many-body localization-like crossover~\cite{BAA2006,OganesyanHuse2007} in finite-size numerical simulation for $W \gg 1$. This regime is not of particular interest to us, as the Hebbian associative memory is essentially destroyed by random fields. 

As we increase the magnon number, states are peeled off the dark band into a single band and eventually the dark band disappears. The ``most strongly correlated'' regime is that at half filling $k=N/2$. For $N=160$ the corresponding Hilbert space size is enormous $n_{\rm Hilbert} \sim 10^{47}$ and is obviously out of reach of any numerical methods. We study an order of magnitude smaller system with $N=16$, $k=8$, $P=3$, and $n_{\rm Hilbert} = 12870$.

At zero field, the density of states is visually reminiscent to Marchenko-Pastur~\cite{MarchenkoPastur1967} and the unfolded level statistics is clear GOE Wigner-Dyson~\cite{BGS1984}. Note that the physical time-reversal symmetry is broken by memories, but an antiunitary symmetry still remains at half filling which is the origin of the orthogonal symmetry class. The application of a field deforms the density of states into a more Gaussian shape~\cite{Kota2001} and drives the level statistics from GOE to GUE. In our finite-size simulations, we observe eventual crossover to Poisson in large fields. But for the fully-connected graph, this MBL-like phenomenon is most certainly an artifact of the finite-size simulations which miss the resonances. It is clear that  the generic quantum $XY$ Hopfield model is quantum chaotic. 

We do not attempt numerical simulations of more complex $SU(d)$ quantum Hopfields, as the exploding Hilbert space size makes most reliable numerical methods impractical. However, it is highly likely that the many-particle spectrum of generic quantum Hopfields is quantum chaotic as well. The exact simulations above teach us an important lesson that the Hebbian memory data is completely lost in random matrix spectra. For example, Figure~\ref{fig:DOS12} ``hides'' just $3$ $XY$-vectors among $12870$ levels. This is a highly impractical way to utilize the Fock space for memory storage. To take advantage of the exponential size of the Hilbert space, one needs to let go of the naive Hebb encoding, find a way to store memories through projectors, and error correct these memories without full measurements. Perhaps, the memory band/dark band separation can provide a way to explore this many-body physics, but we do not attempt it here. Instead, we show that quantum $XY$ and $SU(d)$ Hopfields on fully connected graphs enable physical classical memory recall through mean-field dynamics. 

\begin{figure*}[t]
\centering
\includegraphics[width=0.8\linewidth]{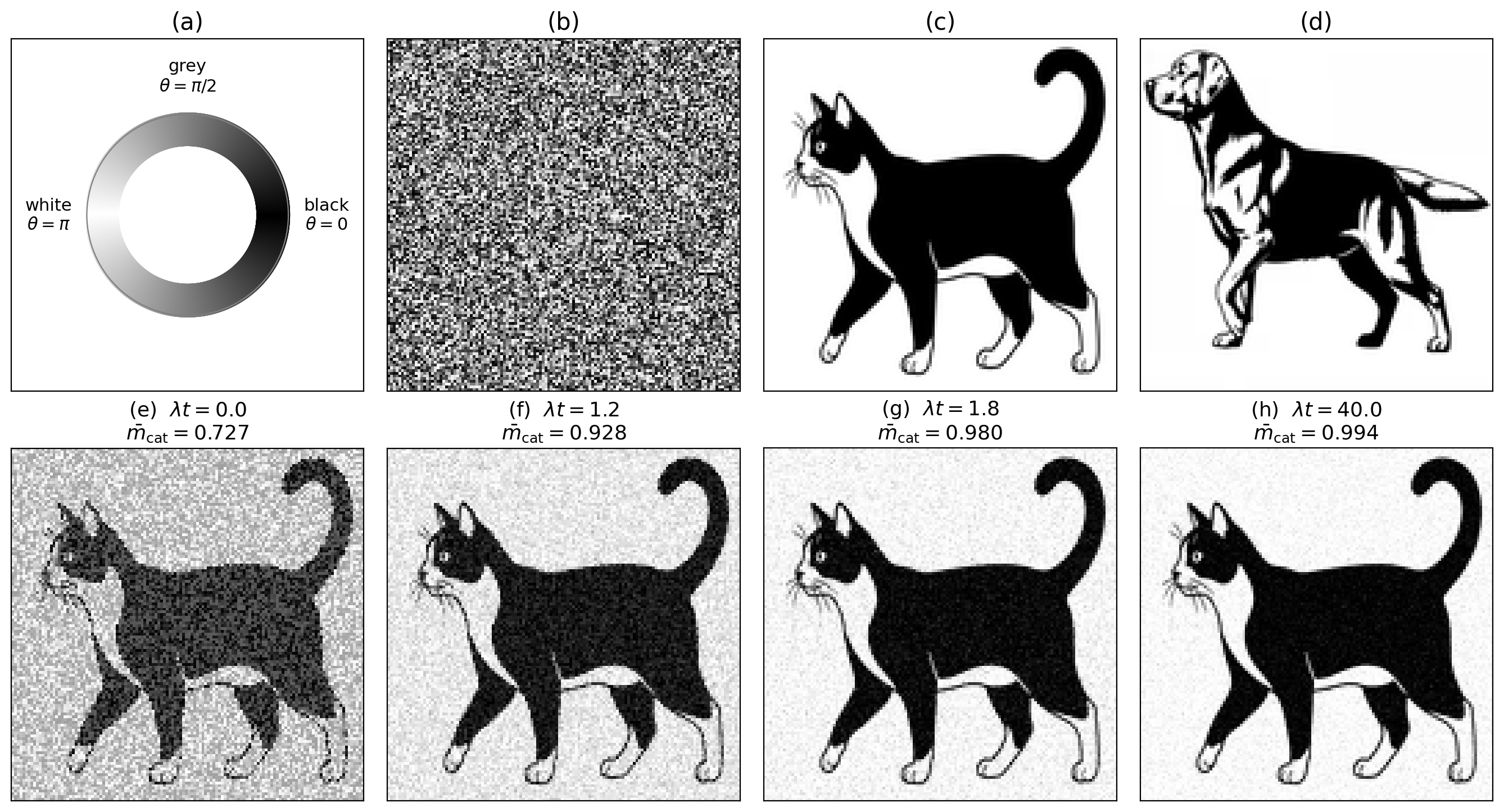}
\caption{(a)~Pixel's luminance $\in [0,1]$
maps to the $U(1)$ phase ${\theta} = \pi \cdot {\rm Luminance}$ of a planar unit spin $e^{i{\theta}}$. E.g., black at $\theta = 0$, grey
at ${\theta} = \pi/2$, white at ${\theta} = \pi$; (b)~An example of a Haar random memory, with pixels' phases drawn independently and
uniformly on $[0, 2\pi)$; (c) and (d)~the $128 \times 128$ patterns corresponding to an image of a cat and dog. Bottom row: The network stores the cat plus {$100$} Haar random patterns and starts with the cue~(e), which is 
the pixelwise blend $\propto 0.52\, e^{i \theta^{\rm cat}} + 0.48\, e^{i \theta^{\rm rand}}$ of the cat with a
stored random pattern. Panels (e)--(h) show the state decoded  along the dissipative
 Landau-Lifshitz-Gilbert (LLG) flow (with $\lambda = 1$), labeled by the time $\lambda t$ in damping units.}
\label{fig:xymem}
\end{figure*}

\section{Memory recall through physical real-time dynamics}
\label{sec:recall}

\subsection{Generalized Landau-Lifshitz-Gilbert equation}

The traditional Hopfield asynchronous memory update rules are algorithmic in the sense that they provide a step-by-step prescription for  a computational process that leads to memory restoration. If we seek to realize/``quantum simulate'' vector and $SU(d)$ Hopfields in physical systems, it is desirable to have a physical process that restores the state of the network to a prescribed memory as a result of natural time evolution. 

We argued that for generalized quantum Hopfields with all-to-all Hebbian interactions - see Eq.~(\ref{Hq}) - mean-field theory works well. In practical terms, it implies that we replace the quantum operators with their expectation values subject to the self-consistency constraints. This essentially brings the model back to the classical formulation we started with. 

\begin{figure*}[t]
\centering
\includegraphics[width=0.95\linewidth]{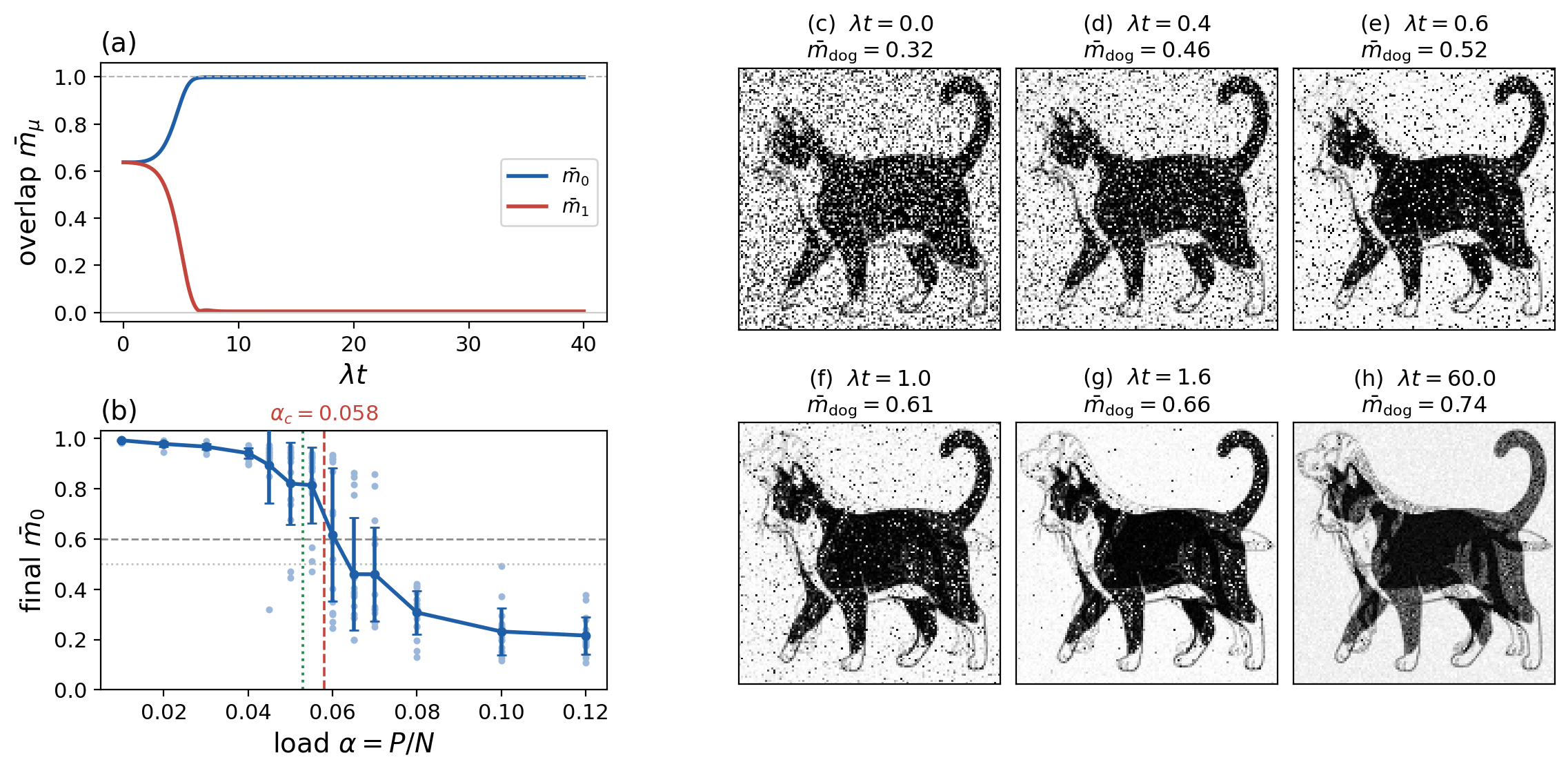}
\caption{(a)~Overlap between the state of the network with two random memories as a function of time of the LLG flow. The initial state is a 50/50 blend of the two memories. The plot demonstrates the winner-takes-it-all phenomenon, where one memory is restored and the other dies out. (b)~ LLG capacity from Haar memories on $\mathbb{S}^2$ ($N=1024$, $R=16$ realizations). The success rate ($\bar{m}_0\ge0.8$) crosses $1/2$ at $\alpha_c\simeq0.058$ (red dashed), slightly exceeding but consistent with the replica $0.053$ estimate (dotted). (c)-(h)~Illustration of the shadow phenomenon. In this $XY$ network, cat, dog, and $100$ random patterns stored ($P=102$). The cue is the cat with $40\%$ of its phases randomized. Even though, the dog image was not cued, the image appears as a subtle shadow during LLG recall in contrast to the winner-takes-it-all phenomenon. This is attributed to the fact that the overlap of the encoded dog/cat pair is $\sim 0.55$, which is well above that for pairs of uncorrelated Haar random memories.}
\label{fig:xyshadow}
\end{figure*}

To formulate the classical equations of motion, we define the generalized vector product through the structure constants $f_{ab}^c$  of the corresponding Lie algebra
$$
\left[ {\bf s} \times_{\rm f} {\bf B} \right]_c \equiv f_{ab}^c s_a B_b,
$$
 For three-dimensional vectors associated with $SU(2)$ this reduces to the familiar cross product. 
Hence, the  mean-field equations of motion of the $i$-th neuron on $\mathbb{C}P^{d-1}$ are
\begin{equation}
\label{Bloch}
\dot{\bf s}_i(t) = {\bf s}_i(t) \times_{\rm f} {\bf B}_i(t)
\end{equation}
The self-consistent field is  produced by the instantaneous
configuration of all other neurons,
\begin{equation}
\label{Bsc}
{\bf B}_i(t) = \sum_{j \ne i} {J}_{ij} {\bf s}_j(t) = \sum_{\mu=1}^{P} {\bm \xi}_i^{\mu}\, O_\mu^{(i)}(t),
\end{equation}
with $O_\mu^{(i)} = \frac{1}{N} \sum_{j \ne i} {\bm \xi}_j^{\mu} \cdot {\bf s}_j$.
This mean-field precession  keeps the classical mean-field energy
${\cal E} \;=\; -\,\frac{1}{2} \sum_{i \ne j} {\bf s}_i \cdot \hat{J}_{ij}\, {\bf s}_j$ intact $\dot{\cal E}=0$ and hence can not possibly lead
to memory recall from a corrupted cue.

To achieve recall without breaking the manifold structure we need a dissipative term that keeps the neuron vectors pinned to it
${\bf s}_i(t) \in \mathbb{C}P^{d-1}$ but ensures that the energy is a non-increasing function of time. 
The minimal structure of this sort is well-known from classical magnetism where it is described
by the Landau-Lifshitz-Gilbert (LLG) equation~\cite{LandauLifshitz1935,Gilbert2004}. The analogous generalized LLG equation reads
\begin{equation}
\label{LLGgen}
\dot{\bf s}_i = {\bf s}_i \times_{\rm f} {\bf B}_i - \lambda\, \left[ {\bf s}_i \times_{\rm f} [{\bf s}_i \times_{\rm f} {\bf B}_i] \right],
\end{equation}
where the last term is the generalized Gilbert damping with $\lambda>0$. Using Eq.~(\ref{LLGgen}) and the mean-field identity
${\bf B}_i = -\,\partial {\cal E}/\partial {\bf s}_i$, it follows that
\begin{equation}
\label{dE}
\dot{\cal E}= - \lambda \sum\limits_i \left|{\bf s}_i \times_{\rm f} {\bf B}_i  \right|^{2}  \le 0.
\end{equation}
Hence the generalized LLG equations ensure that the physical dynamics lowers the energy until a fixed point is reached.
Since the structure of the rugged energy landscape, metastable minima corresponding to memories, and the phase diagram including
the critical capacity are all properties of the energy functional rather than a particular energy-lowering protocol, we expect that 
the LLG dynamics should give rise to efficient memory recall for network loads below critical. The following sections demonstrate it numerically.

\subsection{LLG recall on the Bloch sphere}
We start with the $SU(2)$/Heisenberg Hopfield, but place memories on a circle, ${\bm \xi}_i^\mu = e^{i \theta_i^
\mu} \in \mathbb{S}^1$. {We use them to encode greyscale pixels with $\theta=0$ corresponding to black,
$\theta = \pi/2$ to grey, and $\theta=\pi$ to white, see Fig.~\ref{fig:xymem}(a) for the full legend. We generate $P$ random memories corresponding to $128 \times 128$-pixel square noisy images, see e.g., Fig.~\ref{fig:xymem}(b). In addition we use two meaningful images, ${\bm \xi}_i^{
\rm cat}$ and ${\bm \xi}_i^{\rm dog}$, see Figs.~\ref{fig:xymem}(c) and (d).}

\begin{figure*}[t]
\centering
\includegraphics[width=\linewidth]{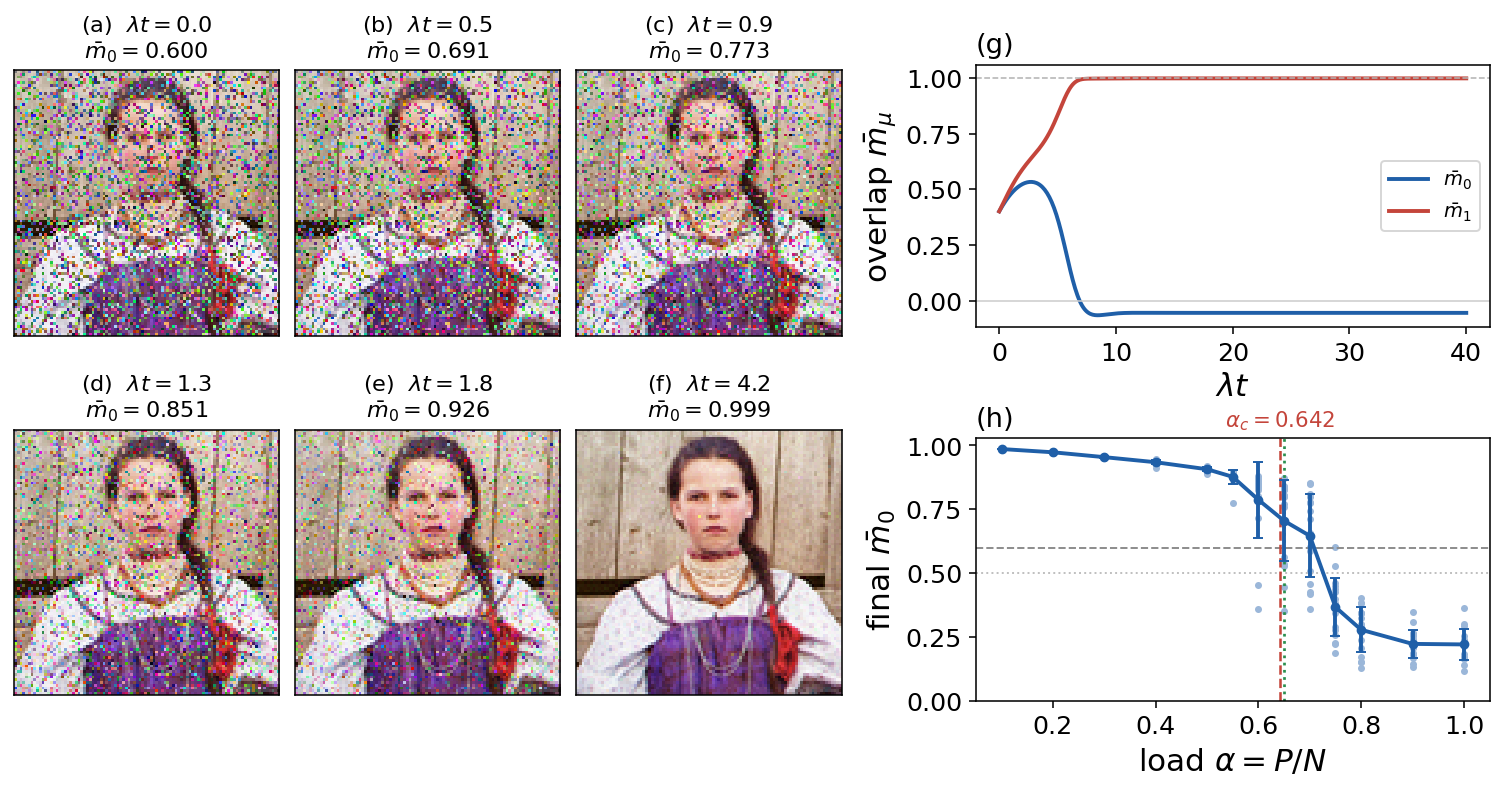}
\caption{(a) A corrupted  photograph with $40\%$ of its $96 \times 96$ pixels replaced by random $SU(3)$ qutrit noise. (b)--(f) Image recall through the generalized Landau-Lifshitz-Gilbert flow on the colored $SU(3)$ Hopfield network. The efficiency of the recall is measured via the overlap $\bar{m}_\mu \;=\; \frac{3}{4}\, O_\mu
\;=\; \frac{1}{2N}\sum_{i=1}^{N} (3\left|\braket{\xi_i^\mu|s_i}\right|^{2}-1)$. Note that it can be negative. For the low load, we observe a rapid restoration of the image to near-perfect match. (g)~Winner-take-all from the exact $50/50$ superposition of two Haar-random $SU(3)$ memories ($N=1024$, $P=10$, $\lambda=1$). The tie seems to be resolved by finite-$N$ fluctuations, where the memory $1$ ``wins'' $\bar{m}_1\to0.998$, while memory $0$ ``loses'' and  settles slightly below zero (we have no explanation for this weak anti-alignment). (h)~$SU(3)$ LLG capacity from Haar qutrit memories on $\mathbb{C}P^2$ ($N=1024$, $R=16$ realizations; $40\%$-randomized cue, $\bar{m}_0\approx0.60$, dashed line): the success rate ($\bar{m}_0\ge0.8$) crosses $1/2$ at $\alpha_c\simeq0.64$ (red dashed line), which is consistent with  the replica estimate $0.62$ (dotted line).}
\label{fig:su3recall}
\end{figure*}

First, to verify that the LLG recall works as expected, we set ${\bm \xi}_i^0 = {\bm \xi}_i^{\rm cat}$ and use $P=100$ random memories with $101$ memories total - ultralow load. The cue is a slightly corrupted ${\bm \xi}_i^{
\rm cat}$ image as follows:
\begin{equation}
\label{blend}
s_j^x(0) + i\, s_j^y(0) \propto 
w\, e^{i\theta^{\rm cat}_j} + (1-w)\, e^{i\theta^{\rm rand}_j},
 s_j^z(0)=0,
\end{equation}
where $j$ is a neuron number and $0 \le w \le 1$ measures the degree to which the chosen image is corrupted. For the demonstration purposes, we choose $w=0.52$. The damped dynamics restores the target within a few damping times, Fig.~\ref{fig:xymem}(e)-(h). Note that the LLG dynamics may tilt the spins out of plane. In this case, we project the spins on the $XY$ plane, extract the relevant angle, and use the legend/convention in Fig.~\ref{fig:xymem} to decode the image. The retrieval overlap of a phasor memory is measured as $\bar{m}_\mu = \frac{1}{N} \left| \sum_j e^{-i\theta_j^\mu} \left( s_j^x + i\, s_j^y \right) \right|$, which equals unity at perfect recall.

Now we increase the network size and scan its load to estimate the critical capacity against the LLG recall procedure. {We find the critical capacity $\alpha_c \approx 0.04$ ($N=4096$), slightly below the previous estimate for the Heisenberg value $\simeq 0.05$. Hence it is the nature of the manifold rather than that of stored memories that dictates the critical capacity.}

Here we also point out two curious effects of the LLG recall. For generic randomly-drawn Haar random memories, we always observe a winner-takes-it-all recall when one memory is recalled and the other drops to near zero; see Figure~\ref{fig:xyshadow}a. However, this is not the case if there are strongly-correlated memories: e.g.,  $\xi^{\rm cat} \sim \xi^{\rm dog}$. If we prepare a cue (\ref{blend}) with no admixture of $\xi^{\rm dog}$, the latter still emerges in recall - we see a shadow of the other image appearing alongside the cued one.

\subsection{LLG recall on the colored Hopfield network}

We now consider the generalized Landau-Lifshitz-Gilbert recall protocol for the
Hopfield network on the symmetric space $\mathbb{C}P^2$. The corresponding $SU(3)$
structure constants form the totally antisymmetric tensor $f_{abc}$ with the nonzero
components
\begin{align}
&f_{123}=1,\qquad
f_{147}=f_{246}=f_{257}=f_{345}=\tfrac{1}{2},\nonumber\\
&f_{156}=f_{367}=-\tfrac{1}{2},\qquad
f_{458}=f_{678}=\tfrac{\sqrt{3}}{2},
\end{align}
together with those obtained from these by antisymmetry of the indices; all components
not related to the above vanish. We solve for the dynamics of the $8$-dimensional
Bloch vectors ${{\bf s}}_i\in\mathbb{R}^8$, which are, however, pinned to a
$4$-dimensional submanifold of $\mathbb{R}^8$. This is because, in addition to the
normalization $s_{i,a}s_{i,a}=\frac{4}{3}$, the $SU(3)$ coherent states are subject to
the non-linear constraint $d_{abc}\,s_{i,b}s_{i,c}=\frac{2}{3}\,s_{i,a}$, where
$d_{abc}$ is the totally symmetric tensor defined by the anticommutator of the
Gell-Mann matrices, $\{\hat\lambda_a,\hat\lambda_b\}=\frac{4}{3}\delta_{ab}\hat{I}+2\,d_{abc}\hat\lambda_c$.
This vector constraint (with no analogue in $SU(2)$, where all $d_{abc}$ vanish)
contains three independent conditions, so that $8-1-3=4=\dim\mathbb{C}P^2$.
The geometry of the symmetric space is respected by the generalized LLG flow equation
which we actually solve,
\begin{equation}
\label{LLGsu3}
\dot{s}_{i,a} = f_{abc}\, s_{i,b}\, B_{i,c}
\;-\;\lambda\, f_{abc} f_{cde}\, s_{i,b}\, s_{i,d}\, B_{i,e},
\end{equation}
where the structure constants are given above, $i=1,2,\dots,N$ labels neurons,
$a=1,2,\ldots,8$, and the mean field $B_{i,a}$ is the self-consistent field of Eq.~(\ref{Bsc}) produced
by the instantaneous configuration of all other neurons,
\begin{equation}
\label{Bsu3}
B_{i,a}(t)=\sum_{j\ne i} J^{ab}_{ij}\, s_{j,b}(t)
=\sum_{\mu=1}^{P}\xi^{\mu}_{i,a}\, O^{(i)}_{\mu}(t),
\end{equation}
with $O^{(i)}_{\mu}=\frac{1}{N}\sum_{j\ne i}\xi^{\mu}_{j,b}\, s_{j,b}$. I.e.\ $B_{i,a}=-\partial{\cal E}/\partial s_{i,a}$. Equivalently,
$B_{i,a}=\Tr\big[\hat\lambda_a \hat{K}_i\big]$: the mean field is the Bloch vector of
the instantaneous memory kernel of Eq.~(\ref{K}), so the stable fixed point of the
damped LLG flow (\ref{LLGsu3}) at each site coincides with the top-eigenvector update rule.

We now use the same $(RGB) \leftrightarrow SU(3)$ encoding/decoding procedure to encode a fragment of the same  image 
from Collection~\cite{ProkudinGorskii}, as in Fig.~\ref{fig:memories}(a). We disturb this image by random colored noise and project it back onto
$\mathbb{C}P^2$ qutrit. Then we run the LLG recall (\ref{LLGsu3}) with $\lambda=1$. The results for low-load recall are shown {in Fig.~\ref{fig:su3recall}(a)-(f): the corrupted photograph ($40\%$ of the pixels randomized) is restored to $\bar{m}_0 = 0.999$ within a few damping times, with the energy decreasing monotonically along the flow.}

{Finally, we repeat the capacity scan under the LLG dynamics with Haar-random memories only, see Fig.~\ref{fig:su3recall}(h). At $N=1024$ with $R=16$ realizations per load, the success-rate criterion gives  $\alpha_c \simeq 0.64$ for $SU(3)$, consistent with the replica value $0.62$ and with the asynchronous eigenvector-update simulations of Sec.~\ref{sec:image}. The physical LLG recall thus saturates the algorithmic capacity of the generalized Hopfield networks.}

{\em Acknowledgements -- }  This work was supported by the US Army Research Office under Grant Number W911NF-23-1-024. This research was partially sponsored by the Secretary of the Air Force Concepts, Development, and Management (SAF/CDM) organization through the SEQCURE2 program at the University of Maryland's Applied Research Laboratory for Intelligence and Security (ARLIS). The author is grateful to Artem Kolesnikov for his help setting up the computational workflow. The author acknowledges discussions with Kartiek Agarwal, Richard Barney, and  Ivar Martin on quantum Hopfield models. This work was performed in part at Aspen Center for Physics, which is supported by National Science Foundation grant PHY-2210452.


\bibliography{refs}

@article{Hopfield1982,
  author  = {Hopfield, J. J.},
  title   = {Neural networks and physical systems with emergent collective computational abilities},
  journal = {Proc. Natl. Acad. Sci. U.S.A.},
  volume  = {79},
  pages   = {2554--2558},
  year    = {1982},
  doi     = {10.1073/pnas.79.8.2554}
}

@article{AGS1985,
  author  = {Amit, Daniel J. and Gutfreund, Hanoch and Sompolinsky, H.},
  title   = {Storing infinite numbers of patterns in a spin-glass model of neural networks},
  journal = {Phys. Rev. Lett.},
  volume  = {55},
  pages   = {1530--1533},
  year    = {1985},
  doi     = {10.1103/PhysRevLett.55.1530}
}

@article{AGS1987,
  author  = {Amit, Daniel J. and Gutfreund, Hanoch and Sompolinsky, H.},
  title   = {Statistical mechanics of neural networks near saturation},
  journal = {Ann. Phys. (N.Y.)},
  volume  = {173},
  pages   = {30--67},
  year    = {1987},
  doi     = {10.1016/0003-4916(87)90092-3}
}

@article{Noest1988,
  author  = {Noest, A. J.},
  title   = {Associative memory in sparse phasor neural networks},
  journal = {Europhys. Lett.},
  volume  = {6},
  pages   = {469--474},
  year    = {1988},
  doi     = {10.1209/0295-5075/6/5/016}
}

@article{Nicoletti2025,
  author        = {Nicoletti, Flavio and D'Amico, Francesco and Negri, Matteo},
  title         = {Statistical mechanics of vector {H}opfield network near and above saturation},
  journal       = {J. Phys. A: Math. Theor.},
  volume        = {58},
  pages         = {505005},
  year          = {2025},
  doi           = {10.1088/1751-8121/ae2bd0},
  eprint        = {2507.02586},
  archivePrefix = {arXiv}
}

@article{Barney2026,
  author        = {Barney, Richard D. and Bhattacharjee, Sharba and Galitski, Victor and Agarwal, Kartiek and Martin, Ivar},
  title         = {Quantum-stabilized patterns in a vector {H}opfield network},
  journal       = {arXiv:2606.06597},
  year          = {2026},
  eprint        = {2606.06597},
  archivePrefix = {arXiv}
}

@article{VG11,
  author        = {Galitski, Victor},
  title         = {Quantum-to-classical correspondence and {H}ubbard-{S}tratonovich dynamical systems: {A} {L}ie-algebraic approach},
  journal       = {Phys. Rev. A},
  volume        = {84},
  pages         = {012118},
  year          = {2011},
  doi           = {10.1103/PhysRevA.84.012118},
  eprint        = {1012.2873},
  archivePrefix = {arXiv}
}

@article{BBP2005,
  author  = {Baik, Jinho and {Ben Arous}, G{\'e}rard and P{\'e}ch{\'e}, Sandrine},
  title   = {Phase transition of the largest eigenvalue for nonnull complex sample covariance matrices},
  journal = {Ann. Probab.},
  volume  = {33},
  pages   = {1643--1697},
  year    = {2005},
  doi     = {10.1214/009117905000000233}
}

@article{EA1975,
  author  = {Edwards, S. F. and Anderson, P. W.},
  title   = {Theory of spin glasses},
  journal = {J. Phys. F: Met. Phys.},
  volume  = {5},
  pages   = {965--974},
  year    = {1975},
  doi     = {10.1088/0305-4608/5/5/017}
}

@article{SK1975,
  author  = {Sherrington, David and Kirkpatrick, Scott},
  title   = {Solvable model of a spin-glass},
  journal = {Phys. Rev. Lett.},
  volume  = {35},
  pages   = {1792--1796},
  year    = {1975},
  doi     = {10.1103/PhysRevLett.35.1792}
}

@article{TAP1977,
  author  = {Thouless, D. J. and Anderson, P. W. and Palmer, R. G.},
  title   = {Solution of `{S}olvable model of a spin glass'},
  journal = {Philos. Mag.},
  volume  = {35},
  pages   = {593--601},
  year    = {1977},
  doi     = {10.1080/14786437708235992}
}

@article{Parisi1979,
  author  = {Parisi, G.},
  title   = {Infinite number of order parameters for spin-glasses},
  journal = {Phys. Rev. Lett.},
  volume  = {43},
  pages   = {1754--1756},
  year    = {1979},
  doi     = {10.1103/PhysRevLett.43.1754}
}

@article{Kanter1988,
  author  = {Kanter, Ido},
  title   = {Potts-glass models of neural networks},
  journal = {Phys. Rev. A},
  volume  = {37},
  pages   = {2739--2742},
  year    = {1988},
  doi     = {10.1103/PhysRevA.37.2739}
}

@article{NoestPRA1988,
  author  = {Noest, Andr{\'e} J.},
  title   = {Discrete-state phasor neural networks},
  journal = {Phys. Rev. A},
  volume  = {38},
  pages   = {2196--2199},
  year    = {1988},
  doi     = {10.1103/PhysRevA.38.2196}
}

@inproceedings{KrotovHopfield2016,
  author        = {Krotov, Dmitry and Hopfield, John J.},
  title         = {Dense associative memory for pattern recognition},
  booktitle     = {Advances in Neural Information Processing Systems 29},
  pages         = {1172--1180},
  year          = {2016},
  eprint        = {1606.01164},
  archivePrefix = {arXiv}
}

@article{Demircigil2017,
  author  = {Demircigil, Mete and Heusel, Judith and L{\"o}we, Matthias and Upgang, Sven and Vermet, Franck},
  title   = {On a model of associative memory with huge storage capacity},
  journal = {J. Stat. Phys.},
  volume  = {168},
  pages   = {288--299},
  year    = {2017},
  doi     = {10.1007/s10955-017-1806-y}
}

@article{MarchenkoPastur1967,
  author  = {Marchenko, V. A. and Pastur, L. A.},
  title   = {Distribution of eigenvalues for some sets of random matrices},
  journal = {Math. USSR-Sb.},
  volume  = {1},
  pages   = {457--483},
  year    = {1967},
  doi     = {10.1070/SM1967v001n04ABEH001994}
}

@article{BGS1984,
  author  = {Bohigas, O. and Giannoni, M. J. and Schmit, C.},
  title   = {Characterization of chaotic quantum spectra and universality of level fluctuation laws},
  journal = {Phys. Rev. Lett.},
  volume  = {52},
  pages   = {1--4},
  year    = {1984},
  doi     = {10.1103/PhysRevLett.52.1}
}

@article{Perelomov1972,
  author  = {Perelomov, A. M.},
  title   = {Coherent states for arbitrary {L}ie group},
  journal = {Commun. Math. Phys.},
  volume  = {26},
  pages   = {222--236},
  year    = {1972},
  doi     = {10.1007/BF01645091}
}

@article{SachdevYe1993,
  author  = {Sachdev, Subir and Ye, Jinwu},
  title   = {Gapless spin-fluid ground state in a random quantum {H}eisenberg magnet},
  journal = {Phys. Rev. Lett.},
  volume  = {70},
  pages   = {3339--3342},
  year    = {1993},
  doi     = {10.1103/PhysRevLett.70.3339}
}

@article{Gorshkov2010,
  author  = {Gorshkov, A. V. and Hermele, M. and Gurarie, V. and Xu, C. and Julienne, P. S. and Ye, J. and Zoller, P. and Demler, E. and Lukin, M. D. and Rey, A. M.},
  title   = {Two-orbital {$SU(N)$} magnetism with ultracold alkaline-earth atoms},
  journal = {Nat. Phys.},
  volume  = {6},
  pages   = {289--295},
  year    = {2010},
  doi     = {10.1038/nphys1535}
}

@article{Ringbauer2022,
  author  = {Ringbauer, Martin and Meth, Michael and Postler, Lukas and Stricker, Roman and Blatt, Rainer and Schindler, Philipp and Monz, Thomas},
  title   = {A universal qudit quantum processor with trapped ions},
  journal = {Nat. Phys.},
  volume  = {18},
  pages   = {1053--1057},
  year    = {2022},
  doi     = {10.1038/s41567-022-01658-0}
}

@article{Gopalakrishnan2011,
  author  = {Gopalakrishnan, Sarang and Lev, Benjamin L. and Goldbart, Paul M.},
  title   = {Frustration and glassiness in spin models with cavity-mediated interactions},
  journal = {Phys. Rev. Lett.},
  volume  = {107},
  pages   = {277201},
  year    = {2011},
  doi     = {10.1103/PhysRevLett.107.277201}
}

@article{Gopalakrishnan2012,
  author  = {Gopalakrishnan, Sarang and Lev, Benjamin L. and Goldbart, Paul M.},
  title   = {Exploring models of associative memory via cavity quantum electrodynamics},
  journal = {Philos. Mag.},
  volume  = {92},
  pages   = {353--361},
  year    = {2012},
  doi     = {10.1080/14786435.2011.637980}
}

@article{RotondoOpen2018,
  author  = {Rotondo, P. and Marcuzzi, M. and Garrahan, J. P. and Lesanovsky, I. and M{\"u}ller, M.},
  title   = {Open quantum generalisation of {H}opfield neural networks},
  journal = {J. Phys. A: Math. Theor.},
  volume  = {51},
  pages   = {115301},
  year    = {2018},
  doi     = {10.1088/1751-8121/aaabcb}
}

@article{Marsh2021,
  author  = {Marsh, Brendan P. and Guo, Yudan and Kroeze, Ronen M. and Gopalakrishnan, Sarang and Ganguli, Surya and Keeling, Jonathan and Lev, Benjamin L.},
  title   = {Enhancing associative memory recall and storage capacity using confocal cavity {QED}},
  journal = {Phys. Rev. X},
  volume  = {11},
  pages   = {021048},
  year    = {2021},
  doi     = {10.1103/PhysRevX.11.021048}
}

@article{Richardson1963,
  author  = {Richardson, R. W.},
  title   = {A restricted class of exact eigenstates of the pairing-force {H}amiltonian},
  journal = {Phys. Lett.},
  volume  = {3},
  pages   = {277--279},
  year    = {1963},
  doi     = {10.1016/0031-9163(63)90259-2}
}

@article{Dukelsky2004,
  author  = {Dukelsky, J. and Pittel, S. and Sierra, G.},
  title   = {Colloquium: {E}xactly solvable {R}ichardson-{G}audin models for many-body quantum systems},
  journal = {Rev. Mod. Phys.},
  volume  = {76},
  pages   = {643--662},
  year    = {2004},
  doi     = {10.1103/RevModPhys.76.643}
}

@article{Galitski2010pairing,
  author        = {Galitski, Victor},
  title         = {Nonperturbative quantum dynamics of the order parameter in the {BCS} pairing model},
  journal       = {Phys. Rev. B},
  volume        = {82},
  pages         = {054511},
  year          = {2010},
  doi           = {10.1103/PhysRevB.82.054511},
  eprint        = {1003.2237},
  archivePrefix = {arXiv}
}

@article{SedrakyanGalitski2010,
  author        = {Sedrakyan, Tigran A. and Galitski, Victor},
  title         = {Boundary {W}ess-{Z}umino-{N}ovikov-{W}itten model from the pairing {H}amiltonian},
  journal       = {Phys. Rev. B},
  volume        = {82},
  pages         = {214502},
  year          = {2010},
  doi           = {10.1103/PhysRevB.82.214502},
  eprint        = {1005.0544},
  archivePrefix = {arXiv}
}

@article{LandauLifshitz1935,
  author  = {Landau, L. D. and Lifshitz, E. M.},
  title   = {On the theory of the dispersion of magnetic permeability in ferromagnetic bodies},
  journal = {Phys. Z. Sowjetunion},
  volume  = {8},
  pages   = {153},
  year    = {1935}
}

@article{Gilbert2004,
  author  = {Gilbert, T. L.},
  title   = {A phenomenological theory of damping in ferromagnetic materials},
  journal = {IEEE Trans. Magn.},
  volume  = {40},
  pages   = {3443--3449},
  year    = {2004},
  doi     = {10.1109/TMAG.2004.836740}
}

@article{BAA2006,
  author  = {Basko, D. M. and Aleiner, I. L. and Altshuler, B. L.},
  title   = {Metal-insulator transition in a weakly interacting many-electron system with localized single-particle states},
  journal = {Ann. Phys. (N.Y.)},
  volume  = {321},
  pages   = {1126--1205},
  year    = {2006},
  doi     = {10.1016/j.aop.2005.11.014}
}

@article{OganesyanHuse2007,
  author  = {Oganesyan, Vadim and Huse, David A.},
  title   = {Localization of interacting fermions at high temperature},
  journal = {Phys. Rev. B},
  volume  = {75},
  pages   = {155111},
  year    = {2007},
  doi     = {10.1103/PhysRevB.75.155111}
}

@article{KropffTreves2005,
  author  = {Kropff, Emilio and Treves, Alessandro},
  title   = {The storage capacity of {P}otts models for semantic memory retrieval},
  journal = {J. Stat. Mech.},
  volume  = {2005},
  pages   = {P08010},
  year    = {2005},
  doi     = {10.1088/1742-5468/2005/08/P08010}
}

@article{Bolle2003spherical,
  author  = {Boll{\'e}, D. and Nieuwenhuizen, Th. M. and P{\'e}rez Castillo, I. and Verbeiren, T.},
  title   = {A spherical {H}opfield model},
  journal = {J. Phys. A: Math. Gen.},
  volume  = {36},
  pages   = {10269--10277},
  year    = {2003},
  doi     = {10.1088/0305-4470/36/41/002}
}

@article{Rebentrost2018,
  author  = {Rebentrost, Patrick and Bromley, Thomas R. and Weedbrook, Christian and Lloyd, Seth},
  title   = {Quantum {H}opfield neural network},
  journal = {Phys. Rev. A},
  volume  = {98},
  pages   = {042308},
  year    = {2018},
  doi     = {10.1103/PhysRevA.98.042308}
}

@article{FiorelliPottsHopfield2022,
  author        = {Fiorelli, Eliana and Lesanovsky, Igor and M{\"u}ller, Markus},
  title         = {Phase diagram of quantum generalized {P}otts-{H}opfield neural networks},
  journal       = {New J. Phys.},
  volume        = {24},
  pages         = {033012},
  year          = {2022},
  doi           = {10.1088/1367-2630/ac5490},
  eprint        = {2109.10140},
  archivePrefix = {arXiv}
}

@article{Bodeker2023,
  author        = {B{\"o}deker, Lukas and Fiorelli, Eliana and M{\"u}ller, Markus},
  title         = {Optimal storage capacity of quantum {H}opfield neural networks},
  journal       = {Phys. Rev. Research},
  volume        = {5},
  pages         = {023074},
  year          = {2023},
  doi           = {10.1103/PhysRevResearch.5.023074},
  eprint        = {2210.07894},
  archivePrefix = {arXiv}
}

@article{Marsh2025,
  author        = {Marsh, Brendan P. and Atri Schuller, David and Ji, Yunpeng and Hunt, Henry S. and Ganguli, Surya and Gopalakrishnan, Sarang and Keeling, Jonathan and Lev, Benjamin L.},
  title         = {High-capacity associative memory in a quantum-optical spin glass},
  journal       = {arXiv:2509.12202},
  year          = {2025},
  eprint        = {2509.12202},
  archivePrefix = {arXiv}
}

@book{Mehta2004,
  author    = {Mehta, Madan Lal},
  title     = {Random Matrices},
  edition   = {3},
  series    = {Pure and Applied Mathematics},
  volume    = {142},
  publisher = {Elsevier/Academic Press},
  address   = {Amsterdam},
  year      = {2004},
  doi       = {10.1016/S0079-8169(04)X8018-1}
}

@article{Kota2001,
  author  = {Kota, V. K. B.},
  title   = {Embedded random matrix ensembles for complexity and chaos in finite interacting particle systems},
  journal = {Phys. Rep.},
  volume  = {347},
  pages   = {223--288},
  year    = {2001},
  doi     = {10.1016/S0370-1573(00)00113-7}
}

@article{Rosenhaus2019,
  author        = {Rosenhaus, Vladimir},
  title         = {An introduction to the {SYK} model},
  journal       = {J. Phys. A: Math. Theor.},
  volume        = {52},
  pages         = {323001},
  year          = {2019},
  doi           = {10.1088/1751-8121/ab2ce1},
  eprint        = {1807.03334},
  archivePrefix = {arXiv}
}

@article{Zhang2014,
  author  = {Zhang, X. and Bishof, M. and Bromley, S. L. and Kraus, C. V. and Safronova, M. S. and Zoller, P. and Rey, A. M. and Ye, J.},
  title   = {Spectroscopic observation of {$SU(N)$}-symmetric interactions in {Sr} orbital magnetism},
  journal = {Science},
  volume  = {345},
  pages   = {1467--1473},
  year    = {2014},
  doi     = {10.1126/science.1254978}
}

@article{Senko2015,
  author  = {Senko, C. and Richerme, P. and Smith, J. and Lee, A. and Cohen, I. and Retzker, A. and Monroe, C.},
  title   = {Realization of a quantum integer-spin chain with controllable interactions},
  journal = {Phys. Rev. X},
  volume  = {5},
  pages   = {021026},
  year    = {2015},
  doi     = {10.1103/PhysRevX.5.021026}
}

@misc{ProkudinGorskii,
  author = {{Library of Congress, Prints and Photographs Division}},
  title  = {The {P}rokudin-{G}orskii Collection},
  year   = {1909},
  url    = {https://www.loc.gov/pictures/collection/prok/}
}

\end{document}